\documentclass[10pt,english,aps,amssymb,prb,twocolumn, showpacs]{revtex4-1}
\pdfoutput=1
\usepackage[utf8]{inputenc}
\usepackage{amsmath}
\usepackage{graphicx}
\usepackage{amssymb}
\usepackage{esint}
\usepackage{verbatim}

\makeatletter
\@ifundefined{textcolor}{}
{
 \definecolor{WHITE}{gray}{1}
 \definecolor{RED}{rgb}{1,0,0}
 \definecolor{GREEN}{rgb}{0,1,0}
 \definecolor{BLUE}{rgb}{0,0,1}
 \definecolor{CYAN}{cmyk}{1,0,0,0}
 \definecolor{MAGENTA}{cmyk}{0,1,0,0}
 \definecolor{YELLOW}{cmyk}{0,0,1,0}
 }

\newcommand{\bra}[1]{\langle #1|}

\newcommand{\brakets}[2]{\left\langle#1| #2 \right\rangle}
\newcommand{\IM}{\text{Im}}
\newcommand{\RE}{\text{Re}}
\renewcommand{\phi}{\varphi}
\renewcommand{\epsilon}{\varepsilon}

\renewcommand{\vec}[1]{{\bf #1}}

\newcommand{\sgn}{\text{sgn}}

\newcommand{\mc}{\mathcal}

\renewcommand{\cite}[1]{[\onlinecite{#1}]}

\usepackage{babel}

\begin{document}
\title {Topological superconductivity and anti-Shiba states in disordered chains of magnetic adatoms}
\author{Alex Westström}
\email[Correspondence to ]{alex.weststrom@aalto.fi}
\author{Kim Pöyhönen}
\author{Teemu Ojanen}
\email[Correspondence to ]{teemuo@boojum.hut.fi}
\affiliation{Department of Applied Physics (LTL), Aalto University, P.~O.~Box 15100,
FI-00076 AALTO, Finland }
\date{\today}
\begin{abstract}
Regular arrays of magnetic atoms on a superconductor provide a promising platform for topological superconductivity. In this work we study effects of disorder in these systems, focusing on vacancies realized by missing magnetic atoms. We develop approaches that allow treatment of ferromagnetic dense chains as well as long-range hopping ferromagnetic and helical Shiba chains at arbitrary subgap energies. Vacancies in magnetic chains play an analogous role to magnetic impurities in a clean $s$-wave superconductor. A single vacancy in a topological chain gives rise to a low-lying "anti-Shiba" state below the band edge of a regular magnetic chain. Proliferation of the anti-Shiba band formed by a finite density of hybridized vacancy states leads to deterioration of the topological phase, which exhibits unusual fragility in a particular parameter region in dilute chains.  We also consider local fluctuation in the Shiba coupling and discuss how vacancy states could contribute to experimental verification of topological superconductivity.         
       
\end{abstract}
\pacs{73.63.Nm,74.50.+r,74.78.Na,74.78.Fk}
\maketitle
\bigskip{}

\section{introduction}
Chains and arrays of magnetic atoms on a superconducting surface offer a promising route to topological superconductivity \cite{qi:2011:1} and accompanying Majorana quasiparticles \cite{kitaev:2001:1}. Different aspects of these systems are under intense experimental \cite{nadj-perge:2014:1,ruby:2015:1,pawlak:2015:1} and theoretical \cite{choy:2011:1,nadj-perge:2013:1,pientka:2013:1,brydon:2015:1,heimes:2014:1,heimes:2015:1,braunecker:2013:1,zhang:2015:1,li:2014:2,rontynen:2015:1,rontynen:2014:1,weststrom:2015:1,klinovaja:2013:1,vazifeh:2013:1,poyhonen:2014:1,nakosai:2013:1,poyhonen:2016:1,rontynen:2016:1,zhang:2016:1} investigation at the moment. While the application potential of magnetic chains seem more rigid compared to the semiconducting nanowire-based realizations \cite{oreg:2010:1,lutchyn:2010:1,mourik:2012:1,das:2012:1,alicea:2011:1} of topological superconductivity, they also offer important advantages over them. Perhaps the most prominent advantage is the fact the magnetic chains can be accurately mapped by Scanning Tunneling Microscopy (STM) techniques.   

The semiconductor nanowire systems with proximity superconductivity are naturally discussed in the language of normal state nanowire properties, such as a Fermi velocity, transverse modes and a mean free path of the wire. These concepts do not have straightforward counterparts in magnetic chains, especially in the dilute Shiba limit. Also, magnetic chains have their own characteristic properties such as long-range hopping between the sites \cite{pientka:2013:1,pientka:2014:1,heimes:2014:1,poyhonen:2016:1,weststrom:2015:1}. Since the early work on the subject \cite{brouwer:2011:1,brouwer:2011:2,stanescu:2011:1,sau:2013:1}, the effects of disorder in the nanowire systems \cite{beenakker:2015:1} and Kitaev's toy model \cite{degottardi:2013:1} have been studied extensively. However, the special properties of magnetic chains have received relatively little attention \cite{hui:2015:1,kim:2014:1}. The purpose of this work is to study the topological and spectral properties of disordered magnetic chains.  Also, since STM techniques enable a single-atom resolved manipulation of the structures, a controlled introduction of defects could be employed to identify the topological phase.

Despite the long history of the subject of magnetic impurities on $s$-wave superconductors, a comprehensive picture has emerged only in the past two decades. Bulk superconductors with arbitrary magnetic impurity concentrations are gapless \cite{balatsky:1997:1} -- in addition to the extended impurity band of hybridized subgap Yu-Shiba-Rusinov states \cite{yu:1965:1,shiba:1968:1,rusinov:1969:1,salkola:1997:1,yazdani:1997:1} there exists rare configurations of lumped impurities. The rare-region fluctuations induce Lifshitz tails to the density of states (DOS), that persist to all subgap energies \cite{balatsky:2006:1}. 
\begin{figure}
\includegraphics[width=0.5\linewidth]{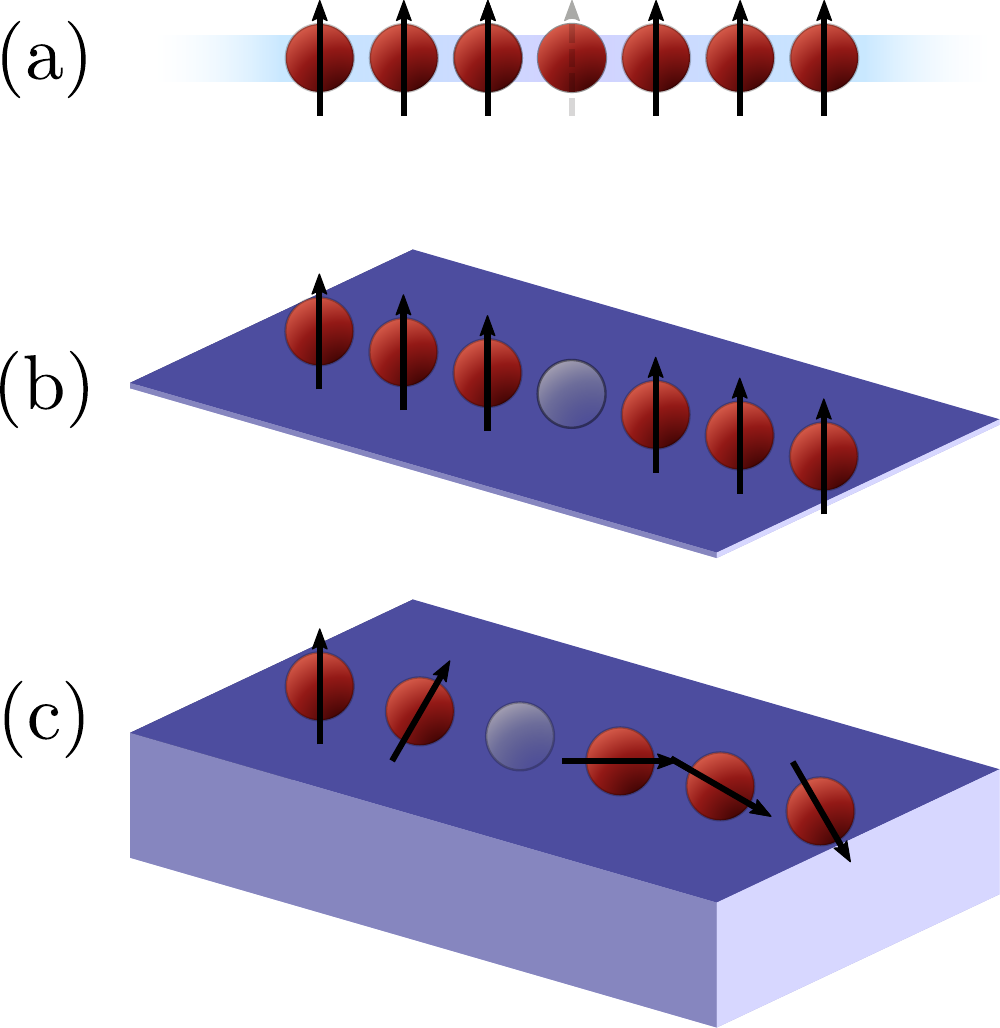}
\caption{Schematic representation of the systems studied in this work, representing (a) the dense chain with nearest-neighbor hopping, (b) the ferromagnetic Shiba chain and (c) the helical Shiba chains. The grayed-out sites represent vacancies giving rise to low-lying anti-Shiba states.}\label{schematic}
\end{figure}
In this work we study vacancies induced by missing magnetic atoms in a regular magnetic chain. This starting point is complementary to the Shiba problem where inhomogeneity is a magnetic impurity atom.  A periodic magnetic atom chain on a superconductor is not disordered -- the subgap bands are characterized by gapped one-dimensional (1D) energy bands. The subgap band may undergo a topological phase transition so that the magnetic lattice forms a topological superconductor. When the magnetic lattice breaks periodicity due to some source of disorder, the system generally acquires low-energy disorder states \cite{montrunich:2001:1,gruzberg:2005:1} that will eventually destroy the topological phase. In addition to vacancies, we consider disorder originating due to locally fluctuating Shiba coupling $\alpha=\pi\nu J S$, where $\nu$ is the DOS of the host superconductor, $J$ is the exchange coupling between the atom and the bulk electrons and $S$ is the impurity spin.  

Below we will treat three different models that exhibit topological superconductivity, illustrated in Fig.~\ref{schematic}. In Sec.~\ref{model} we consider a ferromagnetic short-range hopping model with a Rashba spin-orbit coupling (SOC). This intrinsically 1D model is investigated to illuminate the behavior of densely-packed magnetic chains. In Sec.~\ref{ferro} we consider a complementary parameter regime where the magnetic atoms form a dilute chain and are coupled only through the long-range  hybridization of coupled Shiba wavefunctions. We introduce a formalism which is valid at arbitrary subgap energies and can be employed when the effective low-energy theory does not allow a formulation in terms of a Hamiltonian. In addition to ferromagnetic Shiba chains embedded in a 2D superconductor with a Rashba SOC, in Sec.~\ref{helical} we will analyze  helical Shiba chains embedded in a 3D superconductor.

\section{Dense chain limit}\label{model}
\begin{figure}
\includegraphics[width=\linewidth]{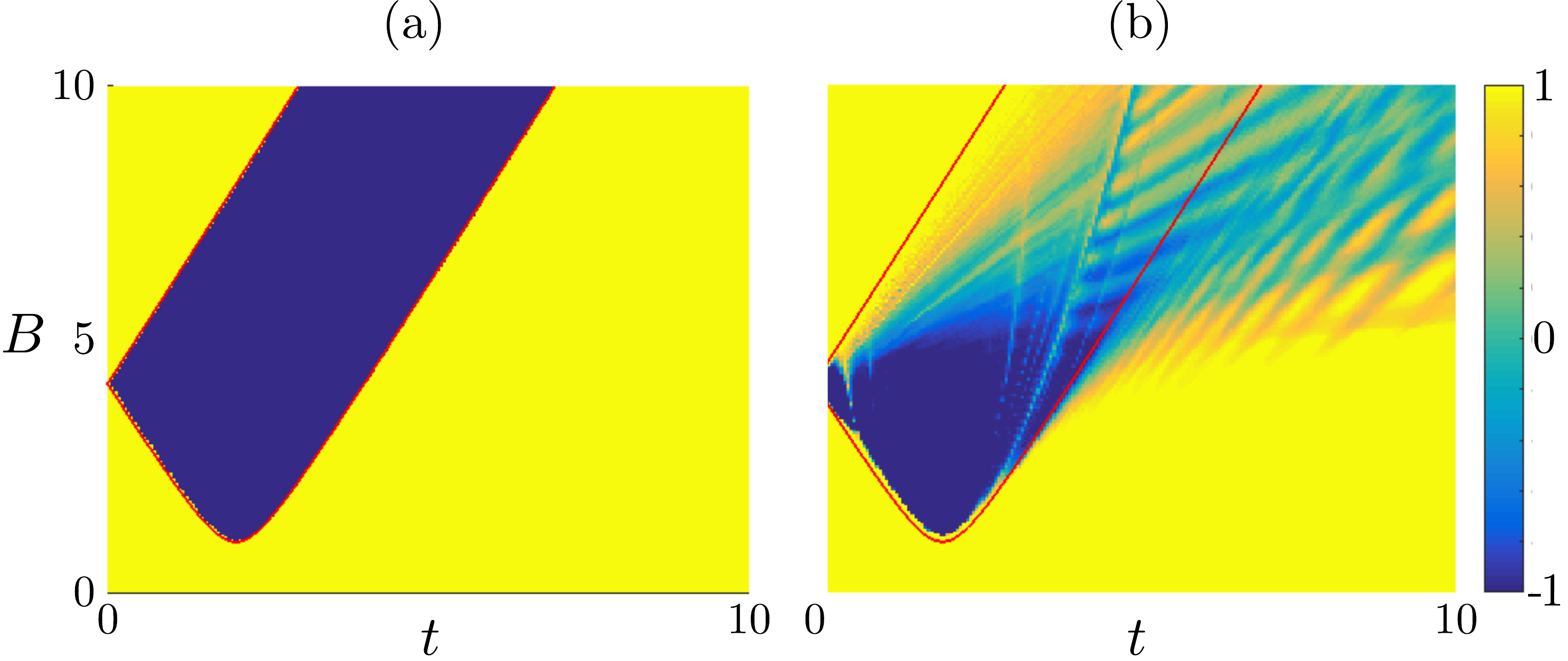}
\caption{Diagram of the $\mathbb Z_2$ invariant of a 50-site dense chain (a) in the pure limit, (b) with 5 vacancies averaged over 200 configurations. Both figures were obtained through applying the Pfaffian inavriant in real space as described in the text. Blue and yellow correspond to the topological and trivial phase, respectively; the red line is the analytical $k$-space solution for the phase boundaries obtained from Eq.~\eqref{eq:dense:topoborder}. Remaining parameters are $\mu = 4$, $\alpha_R = 1$ and $\Delta = 1$.} \label{ferrotwo} 
\end{figure}

We begin by studying a system of a one-dimensional (1D) spin-orbit coupled  superconducting chain which is decorated by magnetic moments. This simple model cannot reproduce all the nuances of realistic systems but has previously revealed much intuition to the phenomenology in magnetic chains. In addition to being more tractable than the microscopically derived long-range hopping models treated in Secs.~\ref{ferro} and \ref{helical}, the model addresses the regime where magnetic moments are packed within a hopping distance apart. In the dense-chain limit, with only nearest-neighbor hoppings considered, the Hamiltonian for this system is
\begin{widetext}
	\begin{equation}\label{eq:dense:H}
		\begin{split}
			H = \sum_{n=1}^{N-1}\sum_{s = \uparrow,\downarrow} (t_n \hat{a}^\dagger_{n+1 s}					\hat{a}_{n s} + t^*_n\hat{a}^\dagger_{n s}\hat{a}_{n+1 					s}) - \sum_{n=1}^N \sum_{s = \uparrow,\downarrow}\mu_n \hat{a}^\dagger_{n s}					\hat{a}_{n s} + \sum_{n=1}^N B_n(\hat{a}^\dagger_{n						\uparrow}\hat{a}_{n\uparrow} - \hat{a}^\dagger_{n						\downarrow}\hat{a}_{n\downarrow})\\ + \alpha_R\sum_{n=1}				^{N-1}\sum_{s_1, s_2 = \uparrow,\downarrow}(\hat{a}^\dagger_{n+1 s_1}(i\sigma_y)_{s_1 s_2}\hat{a}_{n s_2} - \hat{a}^\dagger_{n s_1}(i\sigma_y)_{s_1 s_2}\hat{a}_{n+1 s_2}) + \sum_{n=1}^N(\Delta_n \hat{a}^\dagger_{n\uparrow}\hat{a}^\dagger_{n\downarrow} + \Delta^*_n \hat{a}_{n\downarrow}\hat{a}_{n\uparrow}),
\end{split}
	\end{equation}
\end{widetext}
where $\hat{a}^\dagger_{n s}\ (\hat{a}_{n s})$ creates (destroys) an electron with spin $s$ at site $n$.
In the clean limit, all system parameters are independent of position so that the local hopping amplitudes, chemical potentials, magnetic fields, and superconducting order parameters become $t_n = t$, $\mu_n = \mu$, $B_n = B$ and $\Delta_n = \Delta $, where we with no loss of generality take $\Delta$ to be real. The parameter $\alpha_R$ determines the strength of the Rashba SOC. In this limit, we can formulate a Bogoliubov-de Gennes (BdG) Hamiltonian in $k$-space,
\begin{equation}
	H = \frac{1}{2}\sum_k \Psi^\dagger_k H_k \Psi_k,
\end{equation}
where
\begin{equation}
	H_k = (2t\cos k - \mu) \tau_z + B\sigma_z + 2\alpha_R\sin k \sigma_y \tau_z + \Delta \tau_x
\end{equation}
and $\Psi \equiv (\hat a_{k\uparrow},\hat a_{k\downarrow},\hat a^\dagger_{-k\downarrow}, -\hat a^\dagger_{-k\uparrow})^T$. Here we have set the lattice constant to unity. The matrices $\tau$ and $\sigma$ are Pauli matrices in particle-hole and spin space, respectively. From the above we see that Hamiltonian anticommutes with $\mathcal{C} = \sigma_y\tau_y$ and thus possesses chiral symmetry in addition to the particle-hole symmetry inherent in the BdG formalism.  Hence the system belongs to the symmetry class BDI and supports a $\mathbb{Z}$-valued topological invariant $Q'$ \cite{schnyder:2009:1,ryu:2010:1}. By calculating the winding number one finds $Q' \in \{0,\pm1\}$.  Below we will consider $\mathbb{Z}_2$ phases which are classified by the parity of $Q'$. Following Kitaev \cite{kitaev:2001:1}, we can then for the clean case easily find the borders between the topological and non-topological limits by examining gap closings at $k= 0,\pi$. These occur when
\begin{equation}\label{eq:dense:topoborder}
		B^2 = (2t \pm \mu)^2 + \Delta^2.
\end{equation}
To prepare for the treatment of disordered systems, we will employ a method to evaluate the topological invariant in real space. The $\mathbb{Z}_2$ phases can be identified by studying the response of the ground state fermion parity to twisted boundary conditions. In this case, the $\mathbb{Z}_2$ invariant is given by 
\begin{equation}\label{Q}
Q = \text{sign}\left(\,\text{Pf}[\mc C H_P]\,\text{Pf}[\mc C H_A]\right),		
\end{equation}
where $H_P$ ($H_A$) is the Hamiltonian for the chains with periodic (antiperiodic) boundary conditions and $\text{Pf}[\cdots]$ denotes the Pfaffian of a skew-symmetric matrix. Note that the multiplication of a chiral symmetric Hamiltonian by the corresponding chiral operator $\mc C$ ensures that the resulting matrix is skew-symmetric. The two different boundary conditions can be thought of as a probe for the existence of two different parity sectors achieved by hybridizing the Majorana end modes in open chains \cite{kitaev:2001:1}. The value $Q=1$ corresponds to a trivial state while $Q=-1$ indicates a notrivial state with Majorana end states.  A topological phase diagram for a finite chain in the clean case can be seen in Fig.~\ref{ferrotwo}. Relatively small system of a few dozen of sites will reproduce the infinite system phase diagram essentially perfectly.  The $\mathbb{Z}_2$ phase diagram, shown in Fig.~\ref{ferrotwo} (a), is symmetric with respect to $t\to -t$. As discussed below, disorder will suppress the nontrivial phase of the topological phase diagram of the clean system as depicted in Fig.~\ref{ferrotwo} (b). 

\subsection{Single and two vacancy states}

As in the dual case of dilute magnetic impurities, it is indispensable for understanding the phenomena of multiple defects to first consider a single-defect problem. The defect that we are interested in the present case is a missing magnetic moment, realizing a vacancy in an otherwise perfect lattice of magnetic moments.  We model a system with a single vacancy by setting $B_n$ in Eq.~(\ref{eq:dense:H}) to zero at the site of the defect and everywhere else $B_n = B$. The extension to multiple vacancies should be obvious. The local Zeeman field at a vacancy is suppressed, allowing for bound subgap states.  

\begin{figure}
\includegraphics[width=\linewidth]{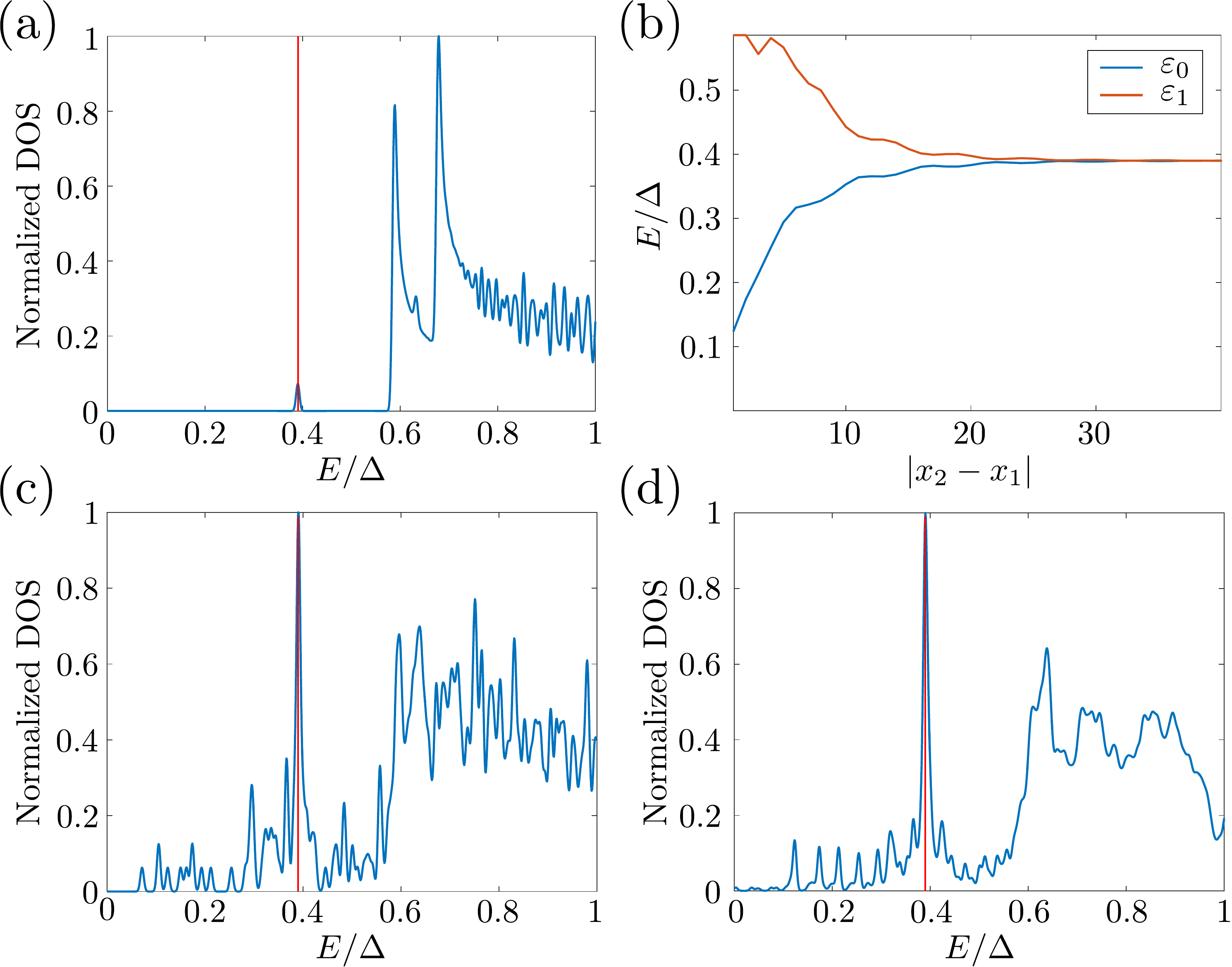} 
\caption{(a) Normalized DOS of a dense chain with PBC, for 1000 sites with a single vacancy. The red line corresponds to the vacancy energy obtained through numerical calculation of the $T$-matrix. (b) Positive subgap bound state energies of a chain with two vacancy sites as a function of the distance between the vacancies. The spectrum has been calculated in a chain with 1000 sites. (c) The DOS of a single vacancy disorder realization of a periodic chain with 2000 sites and 100 vacancies. (d) The DOS for a disorder-averaged ($2\cdot 10^6$ configurations) periodic chain with 80 sites with 4 vacancies. In all four figures, the parameters are $t = 3$, $\mu = 5$, $B = 2$, $\alpha_R = 1$, and $\Delta = 1$.}\label{ferrofour}
\end{figure}

The single-vacancy bound states can be solved by the $T$-matrix formulation \cite{balatsky:2006:1}. Treating Eq.~\eqref{eq:dense:H} as the unperturbed Hamiltonian $H_k$, the $T$-matrix for the system with one vacancy is 
\begin{equation}\label{eq:tmatrix}
T = \left[\mathbb{I}_{4\times4} - \frac{V}{2\pi}\int_{-\pi/a}^{\pi/a} dk G_0(k)\right]^{-1}V,
\end{equation}
where $G_0 = (E-H_k)^{-1}$ is the Green's function of $H_k$, and $V$ is the Fourier transform of the vacancy Hamiltonian. Treating the vacancy as a missing localized magnetic moment, its Fourier transform is simply $-B\sigma_z$, which is independent of $k$, effectively reducing the calculation of the $T$-matrix to an integral over $G_0$. The vacancy energies can then be obtained as poles of the $T$-matrix, or, equivalently, as zeroes of the determinant of the matrix inside the square brackets in Eq.~\eqref{eq:tmatrix}. Due to the involved form of the Green's function of the clean magnetic lattice, the analytical solution is not practical. However, the bound states can be solved numerically as illustrated in Fig.~\ref{ferrofour} (a).  The single-vacancy energy depends on all the parameters. Physically, it is easy to appreciate why the vacancy problem does not allow a simple results such as the single Shiba state: a vacancy in the topological phase can be considered as a hybridized pair of Majorana states localized around the vacancy site. The energy of the Majorana pair will naturally depend on the energy gap of the system, which itself is follows from the complicated dispersion given by Eq.~\eqref{cleanspectrum} in App.~\ref{appclean}.     

Diagonalization of a finite periodic chain with a vacancy site accurately reproduces the subgap states from the $T$-matrix calculation as illustrated in the  Fig.~\ref{ferrofour} (a). Wavefunctions of the vacancy states are localized in the vicinity of the vacancy site.  This is evident in the spectrum of two vacancies as a functions of the distance illustrated in Fig.~\ref{ferrofour} (b).  First, the vacancies are close together, opening up a small trivial region, but as the vacancy sites move apart from each other, the spectrum start to resemble that of two decoupled states. The existence of bound subgap states and their hybridization are the basic building blocks of the low-energy impurity bands that will contaminate the topological state and lead to a gapless phase.

\subsection{Properties of disordered chains}

We now turn  to study disordered configurations with multiple vacancy sites. As argued above, increasing the vacancy density destroys the topological phase by the proliferation of subgap vacancy bands. This behavior is evident in Fig.~\ref{ferrofour} (c) which shows the DOS for a single realization of a long chain and  Fig.~\ref{ferrofour} (d) which illustrates a disorder averaged DOS of shorter chains with 5 \% vacancy concentration. The vacancy band has a large DOS at the single vacancy energy and spreads out to fill the bulk gap. The gap edge of the clean system is smeared out and a finite but suppressed DOS extends all the way down to the gap center.  The DOS of a single long chain  exhibits qualitatively similar features as the disorder averaged DOS with equal vacancy concentration, especially the strong peak at single-vacancy energy.   

It is clear from the Figs.~\ref{ferrofour} (c),(d)  that the spectrum of a topological chain with a robust gap in the clean limit is already dramatically affected by a vacancy concentration of the order a few percent. However, as shown in Fig.~\ref{ferrotwo} (b), the nontrivial phase persists in a significant part of the phase diagram. The disorder-averaged Pfaffian invariant indicates that nontrivial states with low Zeeman splitting are most robust to the vacancy disorder. For the parameters of Fig.~\ref{ferrotwo} (b), the topological phase diagram of disordered system remains by and large unaffected for Zeeman fields $B/\Delta<4$. 

\begin{figure}
\includegraphics[width=0.7\linewidth]{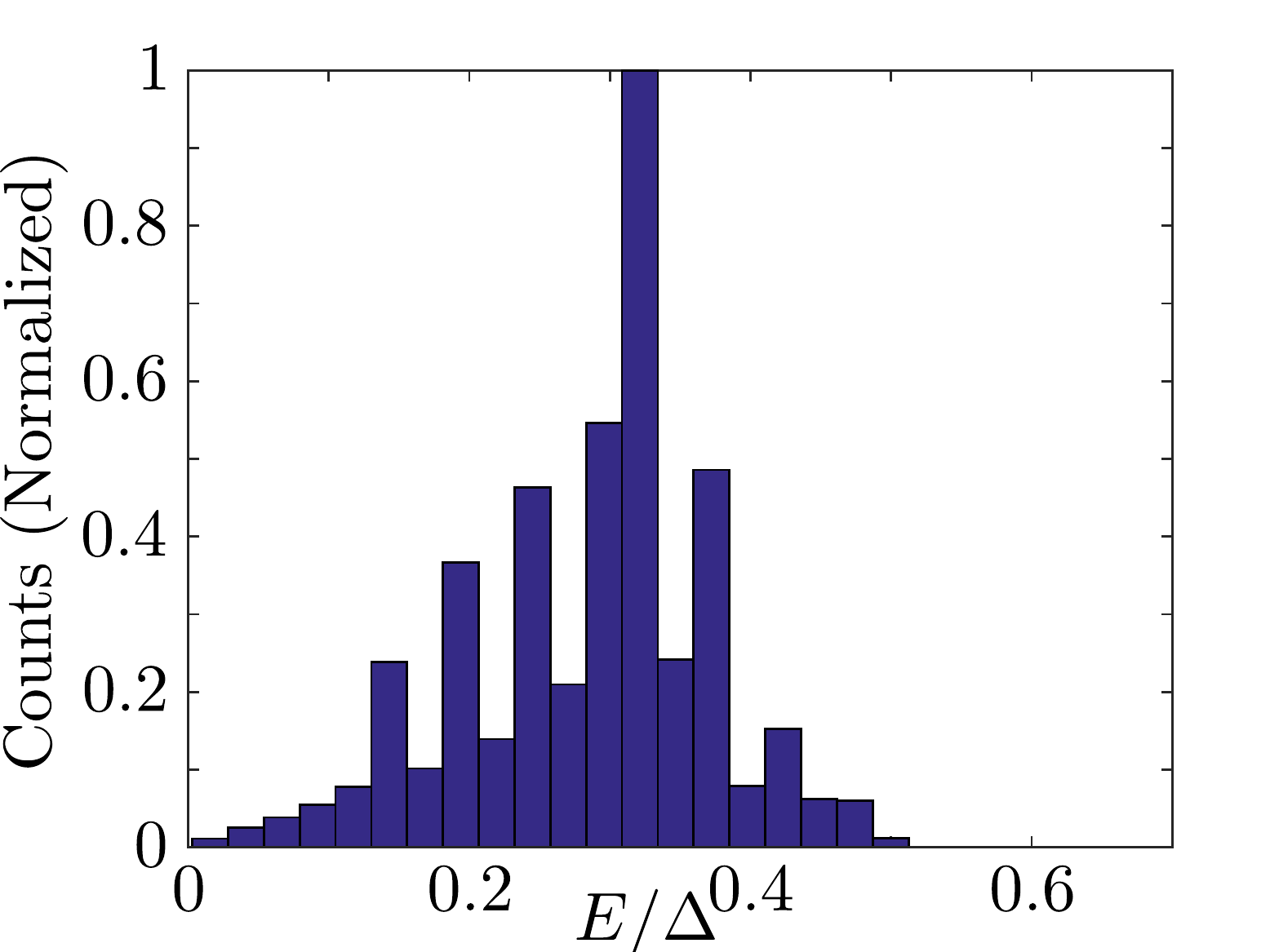}
\caption{Distribution of the lowest-lying positive energy for 50000 realizations of a periodic dense chain with 100 sites of which 10 are vacancies. Parameters are $t = 2.8$, $\mu = 5$, $B = 2$, $\alpha_R = 1$ and $\Delta = 1$. For comparison, the gap energy for a pure system is $ \approx 0.67\Delta$}\label{toygap}
\end{figure}

In quantum information applications of topological superconductors, it is essential that the systems are gapped. The energies of localized Majorana bound states in the vicinity of the gap center should be well-separated from the lowest-lying bulk states. By diagonalizing the system for different configurations, in Fig.~\ref{toygap} we have plotted a distribution of the energy gap for a chain with 100 atoms with 10$\%$ vacancy concentration. Although the gap is significantly suppressed from the clean value, the distribution is peaked at finite energy. The peak structure in the gap distribution reflects the peak structure of the disorder averaged DOS seen in Fig.~\ref{ferrofour} (d). 

At strong disorder topological superconductors are generically expected to exhibit enhanced DOS at zero energy and filling of the excitation gap. In this work we will concentrate on weak or moderate disorder, since magnetic chains are expected to be relatively clean systems. In current experiments the chains are spontaneously formed under suitable conditions. Current technology also allows for the possibility of a top-down fabrication where individual atoms are placed one by one by STM techniques \cite{nilius:2002:1,crommie:1993:1}. In both scenarios it seems likely that high quality can be achieved. Even though effects of weak vacancy disorder in dense chains are not negligible, they are not detrimental to topological properties of finite chains in large parts of the phase diagram.

\section{Ferromagnetic Shiba chain}\label{ferro}
In this section, we consider a model derived from microscopics, consisting of magnetic adatoms deposited on top of a 2D superconducting substrate with Rashba SOC. The system is schematically depicted in Fig.~1 b). The adatoms, treated here as classical spins, form  Yu-Shiba-Rusinov bound states that hybridize to form a band within the gap of the underlying superconductor. Unlike the system considered previously, the starting point of this system is genuinely two-dimensional. In two dimensions, as  observed in a recent experiment \cite{menard:2015:1}, the Shiba states decay as $r^{-1/2}$ at distances up to the order of the coherence length of the underlying superconductor, and exponentially beyond this length scale. As a consequence, the effective low-energy description of a realistic situation where spacing of the magnetic atoms is much smaller than the coherence length may easily involve significant hopping between dozens of closest neighbors. A ferromagnetic chain, in which all spins point in the $z$ direction perpendicular to the plane, is described by the Hamiltonian
\begin{equation}\label{eq:ferroshiba:H}
	\begin{split}
		H = \left(\frac{\vec p^2}{2m}-\mu + \alpha_R(p_y \sigma_x - p_x \sigma_y)\right)\tau_z + \Delta \tau_x \\
		 - J\sum_{i} (\vec S_i \cdot \boldsymbol{\sigma})\delta(\vec r - \vec r_i).
	\end{split}
\end{equation}
The model system has been studied previously in Refs.~\cite{brydon:2015:1,heimes:2015:1,poyhonen:2016:1}. Here we follow the formulation of Ref.~\cite{poyhonen:2016:1} which is valid for the full range of subgap energies. The subgap physics of model \eqref{eq:ferroshiba:H} is conveniently extracted by identifying the relevant subgap degrees of freedom. Each magnetic impurity binds a Shiba state at energies $\pm\Delta\frac{1-\alpha^2}{1+\alpha^2}$ where the dimensionless Shiba coupling $\alpha=\pi\nu J S$ is determined by the exchange coupling $J$, the magnetic moment $S$ and the density of states $\nu$ of the underlying bulk. The long tails of the Shiba states lead to long-range hybrization of states centred at different magnetic moments. The relevant energy scale describing the hybrization of two Shiba states within superconducting coherence length is $\Delta/(k_Fa)^{1/2}$ in 2D and $\Delta/(k_Fa)$ in 3D, where $a$ is the separation of the magnetic moments. 

As explained in App.~\ref{appclean}, we can derive a non-linear eigenvalue problem (NLEVP) for the subgap energy bands, taking the form $\tilde{G}^{-1}(E)\Psi=0$, where
\begin{equation}\label{eq:ferroshiba:NLEVP}
	\tilde{G}^{-1}(E)=\begin{pmatrix}
A\lambda^2-\tfrac{\lambda}{\alpha} & B\lambda & C\lambda^2 & -\lambda D\\
-B\lambda & -A + \tfrac{\lambda}{\alpha} & -\lambda D & C\\
-C\lambda^2 & -\lambda D & A\lambda^2 + \tfrac{\lambda}{\alpha} & -B\lambda\\
-\lambda D & -C & B \lambda & - A -\tfrac{\lambda}{\alpha}
	\end{pmatrix},
\end{equation}
and $\lambda = (\Delta + E)/\sqrt{\Delta^2 - E^2}$. The quantity $\tilde{G}^{-1}$ is related to the Green's function of the chain and is a  nonlinear function of energy, hence the notation and nomenclature.  For a chain consisting of $N$ magnetic moments, $A,\ B,\ C$, and $D$ are $N \times N$ matrices describing the hopping elements between different Shiba states. The detailed expressions for the submatrices can be found in App. \ref{appclean}. Due to the spin and Nambu indices, the spectral problem involves a $4N \times 4N$ matrix. The wavefunctions $\Psi$ have $2N$ electron and $2N$ hole components, containing the information of the spatial localization of the eigenstates along the chain. The matrix elements of the $N \times N$ blocks satisfy the asymptotic behavior $A_{ij}\sim \frac{e^{-a|i-j|/\xi_E}}{|i-j|^{1/2}}$, reflecting the long-range hopping between the Shiba states.  Here $\xi_E = \xi_0/\sqrt{1 - E^2/\Delta^2}$, where $\xi_0 = v_F/\Delta$ is the superconducting coherence length of the underlying superconductor, and $v_F$ is the Fermi velocity of the bulk electrons.
\begin{figure*}
\includegraphics[width=0.8\linewidth]{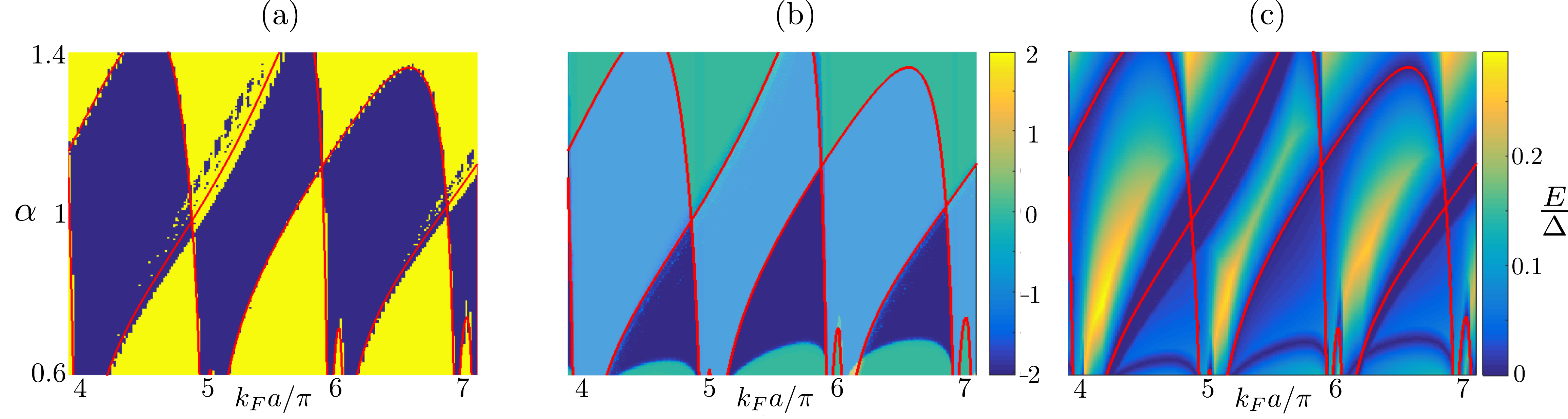} 
\caption{(a) $\mathbb{Z}_2$ phase diagram of the finite ferromagnetic Shiba chain with Rashba SOC, obtained by evaluating the Pfaffian invariant. Blue and yellow correspond to $Q=-1$ and $Q=1$ phases, respectively. The red curves are exact phase boundaries of an infinite system obtained in App.~\ref{appclean}. Parameters used are $\varsigma = 0.01$, $\xi_0 = 50a$ with 50 magnetic sites for the Pfaffian diagram. (b) $\mathbb Z$-valued winding number diagram for an infinite chain with the same parameters. The $|N| = 2$ phase, supporting two Majorana end states in open chains, is not visible in the $\mathbb{Z}_2$ diagrams. (c) Energy gap diagram for the same parameters.}\label{fig:rashbapure}
\end{figure*}

In contrast to the original Hamiltonian in Eq.~\eqref{eq:ferroshiba:H}, the remaining degrees of freedom of the low-energy theory are discrete and localized at the magnetic impurity positions forming a periodic chain with a lattice constant $a$. This effective 1D model, defined by the hopping matrix $\tilde{G}^{-1}(E)$, belongs to the symmetry class BDI, thus supporting a $\mathbb Z$-valued topological invariant. Recently it was discovered that in the physically relevant parameter regime the model supports four different topologically nontrivial phases with one and two Majorana end states \cite{poyhonen:2016:1}. Here we will, however, mostly concern ourselves with the phases of a single Majorana states. This allows us to classify the topological phases with the Pfaffian invariant $Q$ which only distinguishes states with different ground state parities. In the low energy description, the fundamental object is $\tilde{G}^{-1}(E)$  in Eq.~\eqref{eq:ferroshiba:NLEVP} -- not a Hamiltonian. The spectrum of the chain can be computed by solving the equation $\mathrm{det}\left[\tilde{G}^{-1}(E)\right]=0$ for $E$ and finding the eigenvectors belonging to the kernel of $\tilde{G}^{-1}(E)$. Solving NLEVPs is generally resource-consuming compared to linear matrix eigenvalue problems.  However, as explained in App.~\ref{topoH}, the topological properties of the system can be extracted from "topological Hamiltonian" $\tilde{H} =\tilde{G}^{-1}(0)$. 

In Fig.~\ref{fig:rashbapure} (a) we have plotted the $\mathbb Z_2$ topological phase diagram of the system by evaluating the Pfaffian invariant $Q$ for $\tilde{H}$ in a finite chain. This reproduces the clean, infinite system phase diagram accurately. We have reproduced the winding number phase diagram \cite{poyhonen:2016:1} in Fig.~\ref{fig:rashbapure} (b)  for comparison, illustrating the double Majorana phases that are not distinguished by the Pfaffian invariant.  By solving the full NLEVP for a clean system, in Fig.~\ref{fig:rashbapure} (c) we have illustrated the energy gap of an infinite chain. By inspecting the gap diagram it is clear that the small discrepancies between the finite system phase diagram and the analytical phase boundaries of the infinite system in Fig.~\ref{fig:rashbapure} (a) arise only in the parameter regions where the system is nearly gapless and the ground state parity is susceptible to weak perturbations.

\subsection{Topological properties of disordered chains}

\begin{figure}
\includegraphics[width=0.99\linewidth]{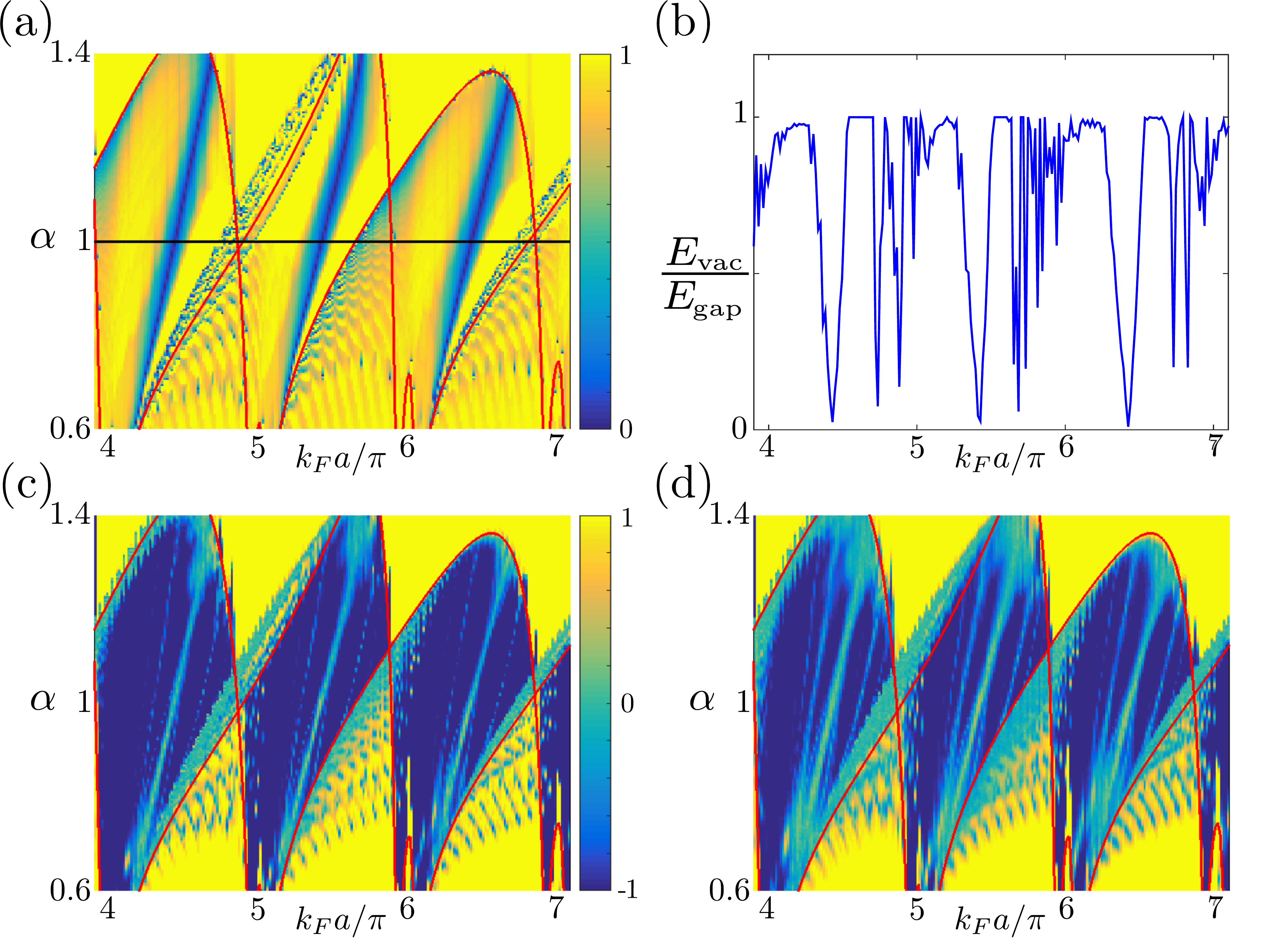}
\caption{Effect of vacancies on the ferromagnetic Shiba chain. (a) Ratio of the single-vacancy bound state energy $E_{\mathrm{vac}}$ to the smallest positive energy $E_{\mathrm{gap}}$ of a clean system. Remaining parameters are $\varsigma = 0.01$, $\xi_0 = 50a$. Note also the diagonal lines where the impurity energy is fine-tuned to near zero. The maximum values have been capped at 1 to avoid divergences at gap closings. (b) Relative gap plotted along the black line in (a). The vacancy-based gap closings are clearly visible, being wider than those caused by topological phase changes. (c) Pfaffian invariant $Q$ for a 50-site system with 5 vacancies, averaged over 400 configurations. The parameters are otherwise the same as in the previous figures. Additional diagonal gapless lines are seen next to the original single-vacancy lines (d) Same, but with 10 vacancies.}\label{rashbafour}
\end{figure}

As in the case of dense chain, we first study the vacancy states due to a single missing magnetic moment. The vacancies are modelled by taking $S_i=0$ in Eq.~(\ref{eq:ferroshiba:H}) at the vacancy sites and $S_i=S$ elsewhere. In the low-energy theory \eqref{eq:ferroshiba:NLEVP}, each vacancy reduces the number of lattice sites by one. A vacancy in the topological phase introduces an anti-Shiba bound state in the band gap of a regular Shiba lattice, as illustrated in Fig.~\ref{rashbafour} (a). In the trivial phase the bound states generally do not exist, while in the nontrivial phase the single vacancy binds a subgap state whenever the gap is robust as seen in  Fig.~\ref{rashbafour} (b). As highlighted in Figs.~\ref{rashbafour} (a) and (b), the single-vacancy energy diagram shows also a striking feature: there is stripe-like pattern inside the nontrivial $\mathbb Z_2$ phase where the vacancy energy lies at the gap center. Furthermore, as a comparison with the gap diagram in Fig.~\ref{fig:rashbapure} (c) reveals,  the vanishing bound state energy is \emph{not} correlated with a small topological gap. At the striped parts of the phase space, the hybridized Majorana pair created by the vacancy has a vanishing excitation energy even when the clean gap is robust.

In Fig.~\ref{rashbafour} (c) and (d) we have plotted topological phase diagrams averaged over different disorder configuration for finite chains with different vacancy concentrations. These illustrate how the topological phase is gradually washed away as the vacancy concentration increases. In the case of multiple vacancies, the bound state energies form a band, the width  of which is determined by the mean distance between the vacancies. It is clear that the vicinity of the phase boundaries of the clean system are fragile since the gap is small and the ground state fermion parity can fluctuate as a result of weak disorder. Even more strikingly, the nontrivial phase of disordered chains is divided by a stripe-like patterns that dissect the clean system's topological phase. This fragility of the topological phase can be qualitatively understood by considering the single-vacancy states. The deterioration of the topological phase is nucleated from the region where the single-vacancy energy vanishes. In those parts even a weak hopping between the impurity sites far apart may push the impurity state below the Fermi energy, switching the ground state parity measured by the Pfaffian invariant. This mechanism allows a nucleation of the trivial state in the middle of the topological phase at very weak vacancy concentrations in finite chains. Increasing the vacancy concentration, the bandwidth of the anti-Shiba band centered at the Fermi energy increases and will drive the proliferation of a gapless state, splitting the nontrivial phase. 

It should be noted that the above discussed effect of finite DOS at the gap center and proliferation of gapless state is distinct from the ubiquitous effect of lumped disorder configurations leading to the Griffiths effect and a peak in DOS at the Fermi energy \cite{montrunich:2001:1,gruzberg:2005:1}. While the Wigner singularity and the Griffiths effect are generic consequences of strong disorder in topological superconductors \cite{brouwer:2011:1,brouwer:2011:2,sau:2013:1,degottardi:2013:1}, the fragility of the phase diagram of finite Shiba chains at weak disorder is caused by the accidental tuning of the single vacancy energy near the gap centre. 

Although we are mostly concentrating on the single Majorana phase defined by value $Q=-1$ of the Pfaffian invariant, the results illustrated above have important consequences on the disordered double Majorana phases indicated in Fig.~\ref{fig:rashbapure} (b). Since $Q$ measures the fermion parity of the ground state, the double Majorana phases with winding numbers $\pm2$ map to the trivial sector $Q=1$. As one can see in Figs.~\ref{rashbafour} (c) and (d), the parity averaged over disorder configurations in a double Majorana phase fluctuates at weak vacancy concentration where most parts of the  $Q=-1$ phases are unaffected. This illustrates that the double Majorana phases are substantially more fragile to disorder than single Majorana phase.    

\begin{figure}
\includegraphics[width=\linewidth]{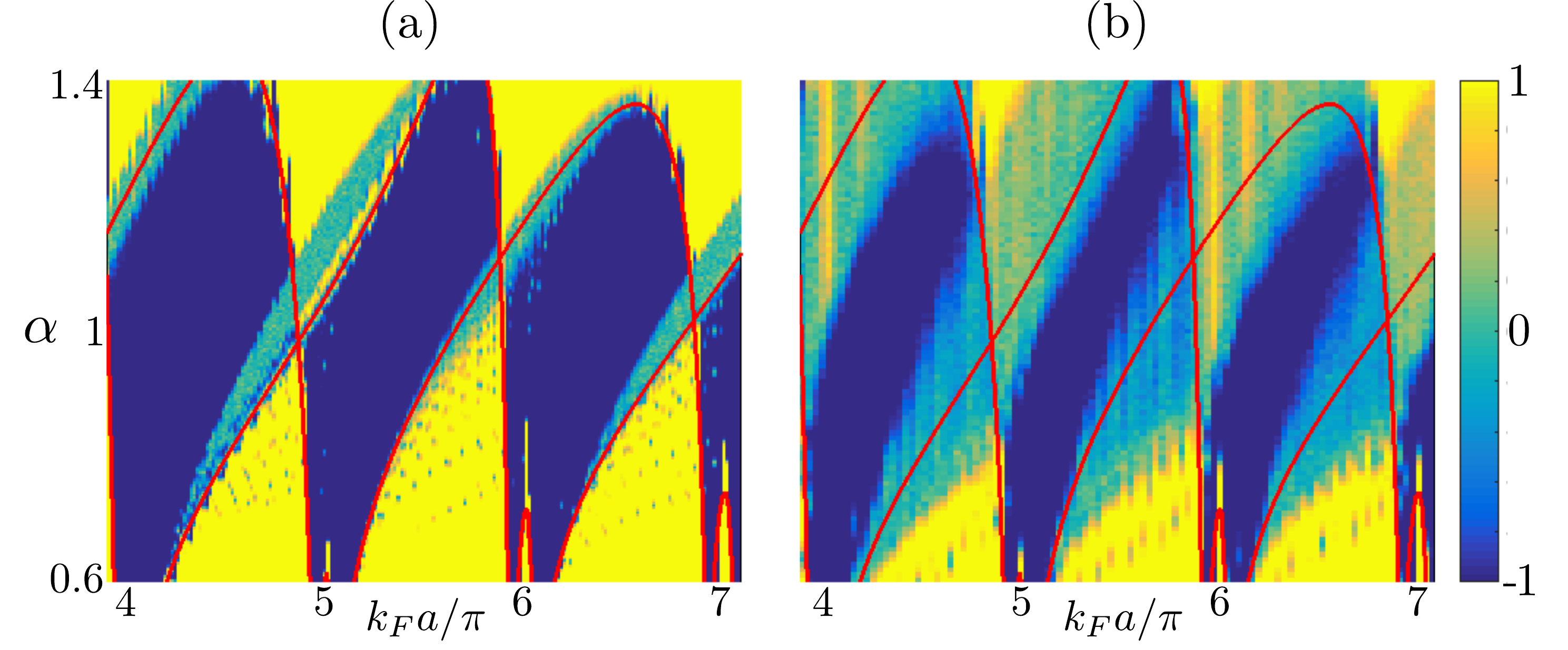}
\caption{The effect of disorder in the Shiba coupling $\alpha$ on the topology of the ferromagnetic Shiba chain. An average over (a) 100 configurations for a uniform disorder of 5\% in $\alpha$; (b) 200 configurations for 20\% disorder. Other parameters are $\varsigma = 0.01$, $\xi_0 = 50a$. The length of the PBC chain is 50 sites.}\label{rashba_alpha}
\end{figure}
In addition to studying vacancies, we briefly consider an onsite disorder in the dimensionless Shiba coupling $\alpha=\pi\nu J S$ on the topology of the system.  Recently it was suggested that disorder in the underlying superconductor may lead to variation in this parameter \cite{hui:2015:1}. In addition,  variation of the exchange coupling $J$, originating from different microscopic coupling configurations between the atom and the substrate, will also translate into variation of $\alpha$. Recently this was employed in the observation of the ground state parity switching that takes place in a system with an isolated magnetic impurity at $\alpha=1$ \cite{hatter:2015:1}. We will allow $\alpha$ to vary locally as $\alpha_i = \alpha(1+ \delta\alpha_i)$, where $\delta\alpha$ is a uniformly distributed with a finite bandwidth $\delta\alpha_i \in [-\epsilon,\epsilon]$. The modification to the low-energy theory (\ref{eq:ferroshiba:NLEVP}) due to a fluctuating $\alpha$ is explained in App.~\ref{topoH}. As seen in Figs.~\ref{rashba_alpha} (a) and (b), the system is very robust against this type of disorder. Even at strong disorder when the site-to-site fluctuation of $\alpha$ can reach 20\%, the  nontrivial $\mathbb Z_2$ regions persist for the most parts. As expected, the nontrivial regions with a small energy gap are first washed away. Contrary to the vacancy disorder, the nontrivial phase is deteriorating only from the boundaries without splitting to additional disconnected parts. The double Majorana phases are again completely smeared out at disorder strengths where most parts  of the $Q=-1$ phase still persist.

\subsection{Energy gaps of disordered Shiba chains}

In the previous section we analyzed the $\mathbb Z_2$ topological invariant in the ferromagnetic Shiba chain with disorder. Here we study how the excitation gaps of these systems are affected by vacancy disorder. 

\begin{figure}
\includegraphics[width=0.8\linewidth]{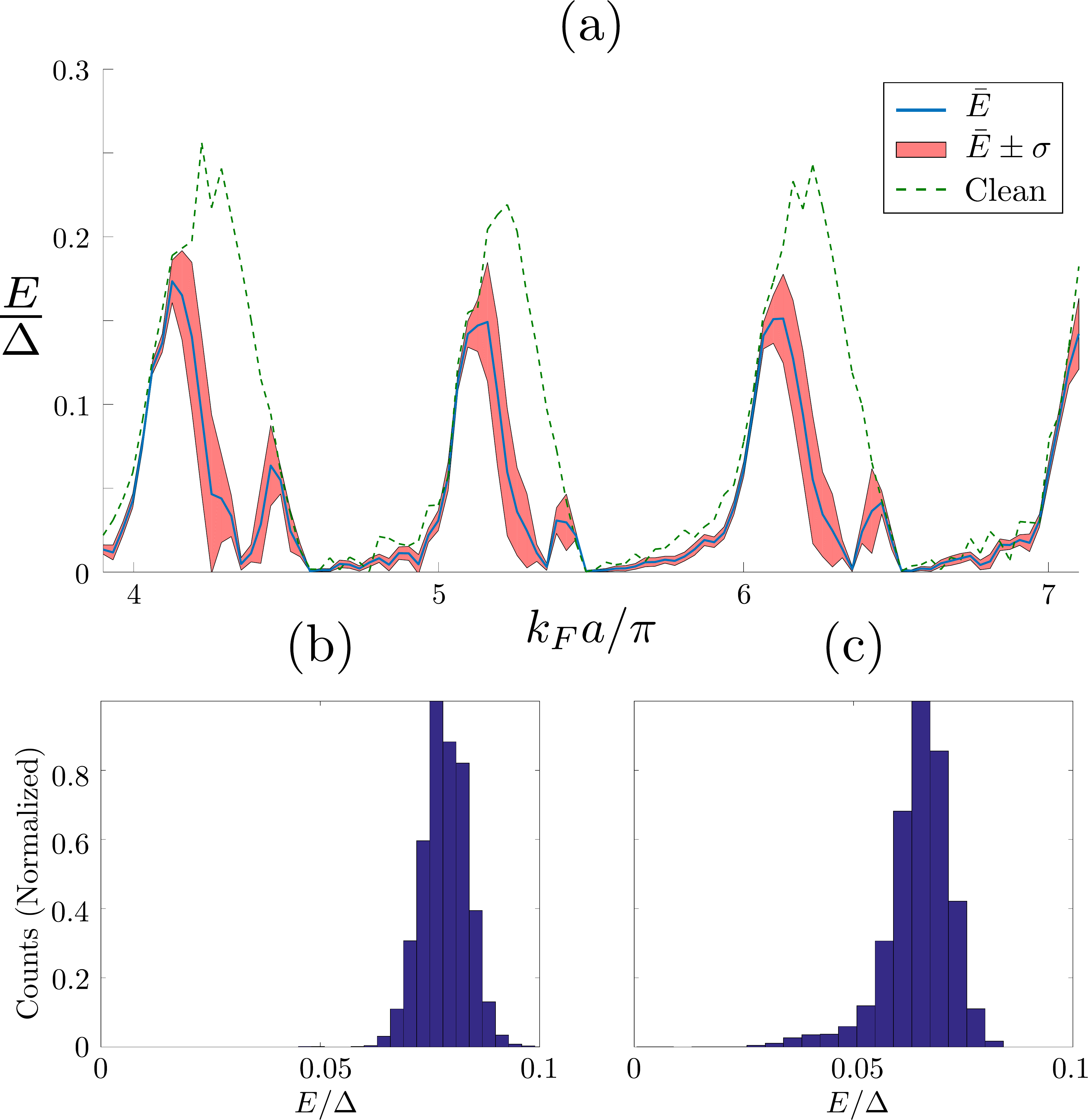}
\caption{(a) Disorder-averaged energy gap $\bar{E}$ for a 100-site ferromagnetic Shiba chain with 10 vacancies, averaged over 1000 configurations. The tube containing the average value represents one standard deviation. Other parameters are $\alpha = 0.9$, $\varsigma = 0.01$, $\xi_0 = 50a$.
(b) and (c): Distribution of the lowest positive energy for a 50-site periodic ferromagnetic Shiba chain with 5 and 10 vacancies, respectively.  Distribution is for 10000 configurations. Other parameters are $k_Fa = 20.5$, $\alpha = 1$, $\varsigma = 0.01$, $\xi_0 = 50a$. The gap for a clean system is $\approx 0.11\Delta$.}\label{rashba_gap}
\end{figure}

In Fig.~\ref{rashba_gap} (a) we have plotted a disorder averaged energy gap of a finite chain. This information should be contrasted to the $\alpha=0.9$ cut of the phase diagram depicted on Fig.~\ref{rashbafour} (c). In certain parts of the nontrivial phase the gap follows very closely to the clean system value. However, as the stripe pattern of the vanishing single-vacancy energy is approached, the gap is dramatically suppressed. As Fig.~\ref{rashba_gap} a) clearly shows, the gap or disordered system closes at the stripe while it approaches the maximum value at the clean system. As argued above, this unexpected fragility of the nontrivial phase is present even for very weak disorder. The statistical fluctuations of the energy gap are pronounced at the left side of the stripe where the average gap closes. This effect shows up in Figs.~\ref{rashbafour} (c) and (d) as the side bands of the main stripe of the vanishing single-vacancy energy.  

In Figs.~\ref{rashba_gap} (b) and (c) we have plotted the gap distributions for two vacancy concentrations. Although both distributions are peaked at energies below the clean system gap $\sim 0.1\Delta$, the topological gap still remains in the observable range. As expected, the distribution for the higher vacancy concentration is peaked and exhibits a tail to lower energies.

\section{Helical Shiba chains }\label{helical}

Finally we analyze the helical Shiba chain. Similarly to the ferromagnetic Shiba system, it consists of classical spins placed in a dilute chain on a superconducting substrate. However, now the substrate is treated as a genuinely three-dimensional bulk. The main differences to the 2D ferromagnetic model are the fact that the Shiba states decay as $r^{-1}$ (instead of $r^{-1/2}$) at distances up to the order of the coherence length and that no Rashba coupling is required to achieve a topologically nontrivial phase \cite{nadj-perge:2013:1}. The Hamiltonian for a helical chain is
\begin{equation}\label{eq:helical:H}
		H = \left(\frac{\vec p^2}{2m}-\mu \right)\tau_z + \Delta \tau_x 
		 - J\sum_{i} (\vec S_i \cdot \boldsymbol{\sigma})\delta(\vec r - \vec r_i).
\end{equation}
The magnetic moments form a helical texture, $\mathbf{S}_j = (\sin(\theta)\cos(\phi_j),\sin(\theta)\sin(\phi_j),\cos(\theta))$ where, for a chain aligned along the $x$ axis, $\phi_j = 2k_Hx_j=2jk_Ha$. The helical wavenumber $k_H$ determines the pitch of the helix. The helical Shiba chain has been examined in more detail in Refs.~\cite{pientka:2013:1,pientka:2014:1,weststrom:2015:1}. Following the treatment in Ref.~\cite{weststrom:2015:1} and also explained in App.~\ref{appclean}, we can again formulate the subgap spectral problem of $N$ magnetic atoms as $\tilde{G}^{-1}(E)\Psi=0$, where
\begin{equation}\label{eq:helical:NLEVP}
	\begin{split}		
		\tilde{G}^{-1}=\begin{pmatrix}
			\lambda^2h^{\uparrow\uparrow}-\frac{\lambda}{\alpha} & -\lambda d^{\uparrow\downarrow} & -\lambda^2h^{\uparrow\downarrow} & \lambda d^{\uparrow\uparrow}\\
			-\lambda d^{\downarrow\uparrow} & -h^{\downarrow\downarrow}+ \frac{\lambda}{\alpha} & \lambda d^{\downarrow\downarrow} & h^{\downarrow\uparrow}\\
			-\lambda^2h^{\downarrow\uparrow} & \lambda d^{\downarrow\downarrow} & \lambda^2h^{\downarrow\downarrow}+ \frac{\lambda}{\alpha} & -\lambda d^{\downarrow\uparrow}\\
			\lambda d^{\uparrow\uparrow} & h^{\uparrow\downarrow} & -\lambda d^{\uparrow\downarrow} & -h^{\uparrow\uparrow} - \frac{\lambda}{\alpha}\\
		\end{pmatrix}
	.
	\end{split}
\end{equation}
The explicit expressions for the $N\times N$ submatrices $h^{\sigma\sigma'}$ and $d^{\sigma\sigma'}$, describing a long-range hopping with asymptotic behavior $h_{ij}^{\sigma\sigma'}, d^{\sigma\sigma'}_{ij} \sim \frac{e^{-a|i-j|/\xi_E}}{|i-j|}$, are given in App. \ref{appclean}. The general helical chain is in symmetry class D; however, the planar helical chain ($\theta = \pi/2$) is in class BDI \cite{poyhonen:2014:1}. For simplicity, we will here  focus on the $\mathbb Z_2$ topological phase of the planar chain, as the $\mathbb Z_2$ boundaries are independent of $\theta$. Analogously to the ferromagnetic chain, we can then again define a topological Hamiltonian $\tilde H=\tilde{G}^{-1}(0)$ from which the topological properties of the system can be obtained. 

\begin{figure}
\includegraphics[width=\linewidth]{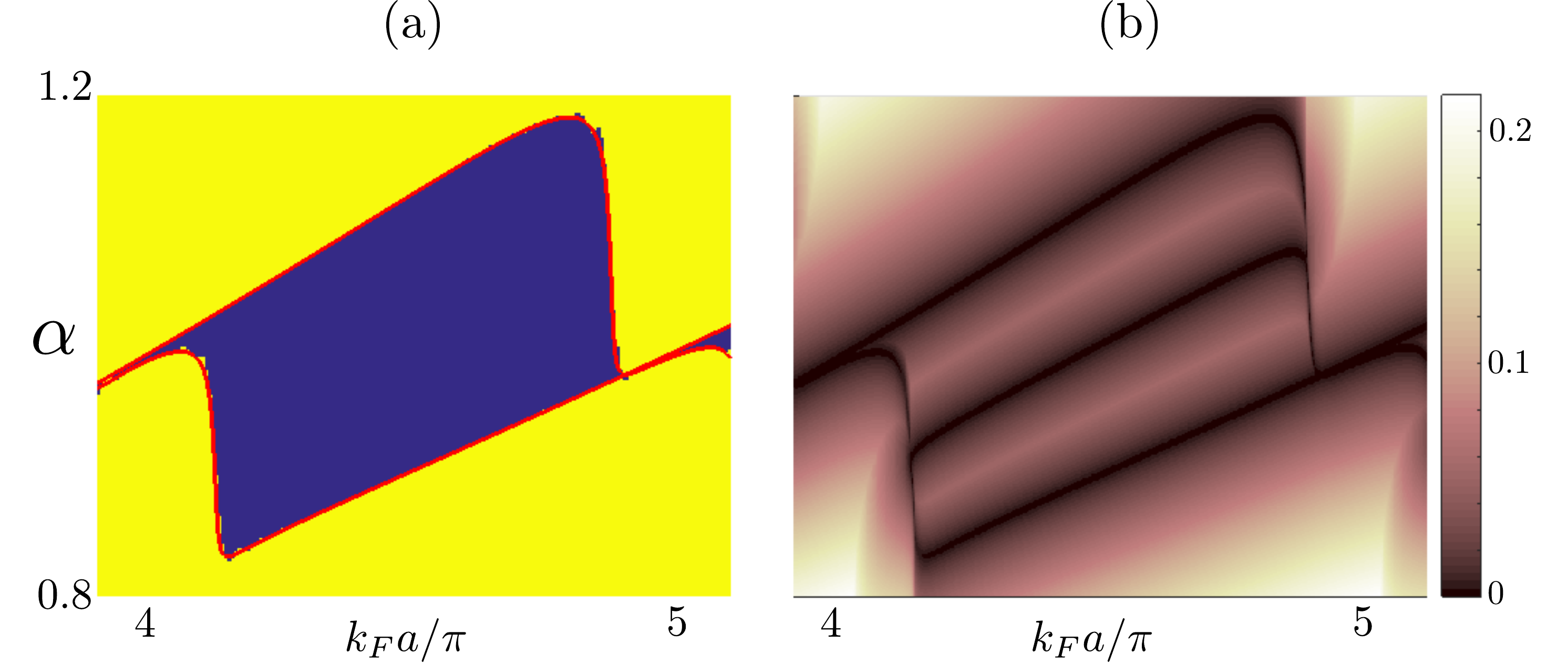} 
\caption{(a) $\mathbb{Z}_2$ phase diagram of a helical Shiba chain of 100 sites with $k_Ha = \pi/8$, $\theta = \pi/2$, $\xi_0 = 50a$. Blue and yellow correspond to values $Q=-1$ and $Q=1$ of the Pfaffian invariant. The red curves represent exact phase boundaries of infinite system, obtained in App.~\ref{appclean}. (b) Energy gap diagram of an infinite helical Shiba chain with the same parameters. The diagonal line visible in the middle here, but not in the $\mathbb Z_2$ diagram, corresponds to the phase transition between phases $N = 1$ and $N = -1$ of $\mathbb Z$-valued winding number invariants.}\label{purehelix}
\end{figure}

By evaluating Pfaffian invariant $Q$  for a finite chain, we have plotted the $\mathbb Z_2$ phase diagram of the pure system in Fig.~\ref{purehelix} (a). Again, $Q$ evaluated for a finite chain reproduces very accurately the infinite system phase diagram.  The energy gap for an infinite system can be achieved by solving the full nonlinear problem. The gap, illustrated  in Fig.~\ref{purehelix} (b), reveals that the nontrivial phase is divided into two disjoint regions by a gap-closing line.

\subsection{Disordered helical chains}

\begin{figure}
\includegraphics[width=\linewidth]{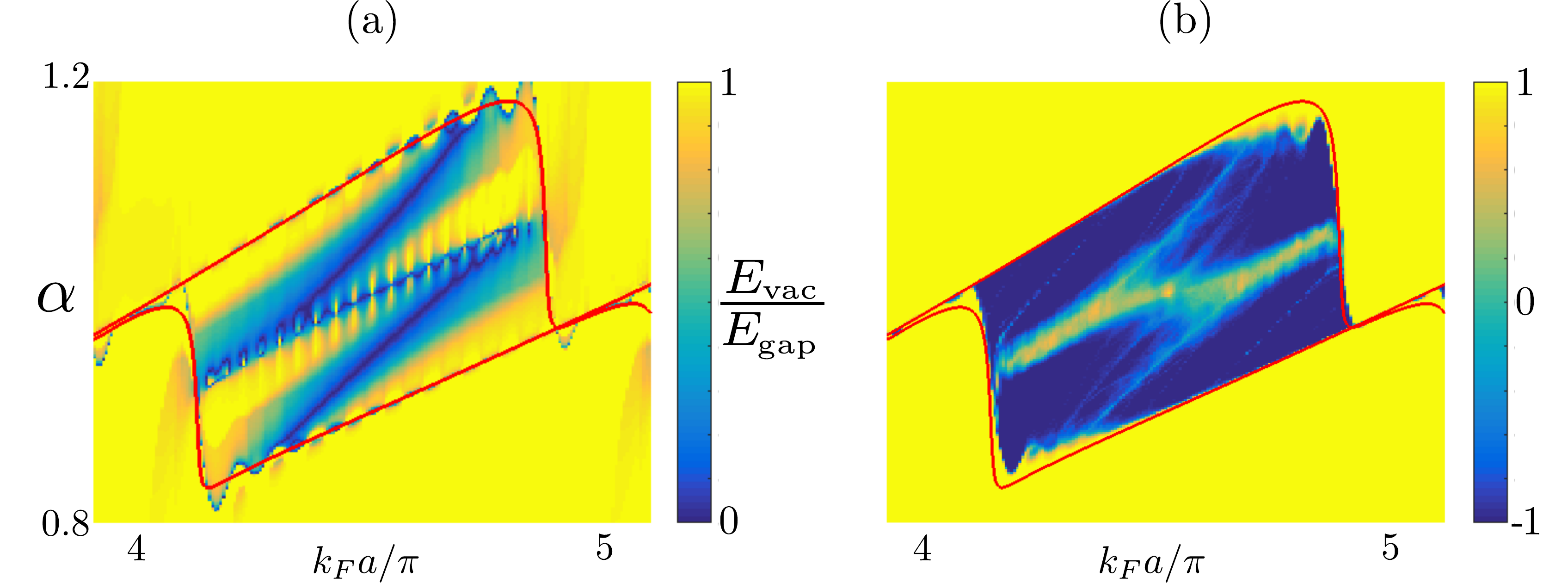} 
\caption{(a) Single-vacancy bound state energy divided by the energy gap of a clean system. Similarly to the ferromagnetic case, a line with zero bound-state energy is observed. The figure is calculated for a chain with 48 sites and $k_Ha = \pi / 8$, $\theta = \pi/2$, $\xi_0 = 50a$. (b) Topological $\mathbb Z_2$ invariant for a helical Shiba chain with 48 sites of which 6 are vacancies, averaged over 400 configurations. Otherwise, the parameters are the same as in (a).}\label{helixdiss}
\end{figure}

We assume that vacancies in the helical chain do not affect the spin configuration of the other sites, so that we can model vacancies by simply removing magnetic sites from a static texture.  Similarly to those found in the ferromagnetic Shiba chain, the single-vacancy bound states are always present in the nontrivial phase. Importantly, we again uncover stripe-like features where the single-vacancy energies lie at the gap center as illustrated in Fig.~\ref{helixdiss} (a). As in the ferromagnetic chain, the vanishing vacancy energy stripes may exist in regions that have robust topological gaps, splitting the nontrivial phase into several disconnected pieces. The stripe patterns, together with the gap-closing line splitting the nontrivial phase of the clean system depicted in Fig.~\ref{purehelix}, are expected to be fragile regions of the nontrivial phase in the presence of disorder. This expectation is confirmed in Fig.~\ref{helixdiss} (b) which shows the typical behavior of the phase diagram averaged over different configurations of multiple vacancies. The nucleation of a gapless phase in the vicinity of fragile regions may take place already at weak disorder. 

\begin{figure}
\includegraphics[width=\linewidth]{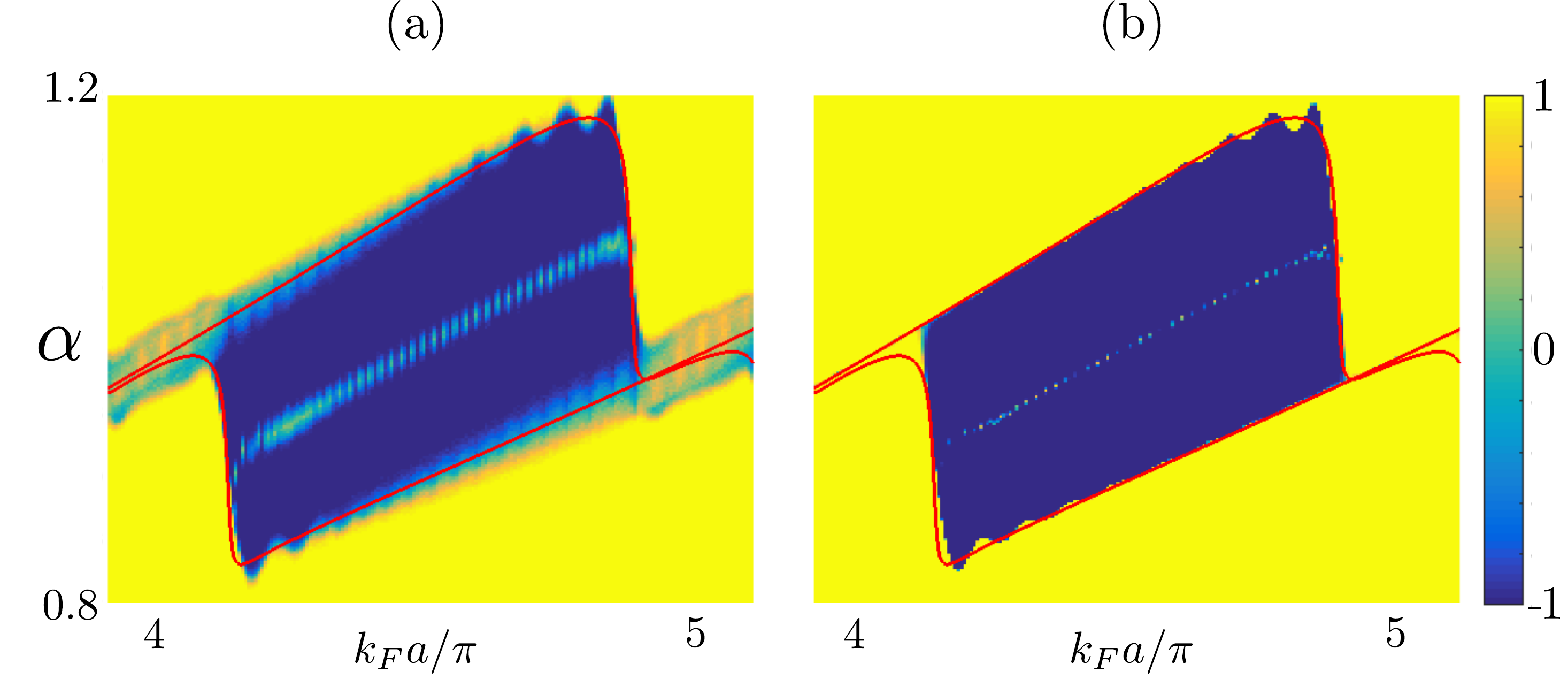}
\caption{Topological $\mathbb{Z}_2$ phase diagrams for a helical Shiba chain with 48 sites and (a) 5\% disorder in $\alpha$, averaged over 300 configurations; (b) Disorder in the planar angles of the magnetic moments: $\phi_i \to \phi_i \pm \delta\phi_i$, where $\delta\phi_i \in [-\pi/8,\pi/8]$, averaged over 100 configurations. Parameters used are $k_Ha = \pi/8$, $\theta = \pi/2$, $\xi_0 = 50a$. } \label{alphahelix}
\end{figure}

As we did for the ferromagnetic chain, we also consider a local random variation in the Shiba coupling $\alpha$. In Fig.~\ref{alphahelix} (a), we see how the disorder averaged $\mathbb Z_2$ invariant behaves. Here we notice that $\alpha$ disorder leads to a diminished nontrivial phase through proliferation of the gapless phase nucleated at the phase boundaries of the clean system. However, the stripe patterns associated with vacancies are absent. As in the ferromagnetic chain,  $\alpha$ disorder in helical chains does not provide big surprises.

We also studied onsite disorder in the pitch angle of the magnetic helix. It turns out, as depicted in Fig.~ \ref{alphahelix} b), that the system is generally highly robust against this type of disorder. As the figure illustrates,  a strong onsite random variation in the magnetization direction comparable to the pitch angle $k_Ha$ of the helix causes a slight modulation of the phase boundaries but leaves the nontrivial phase otherwise intact. 

\section{Discussion and conclusions}

In this work we have studied effects of vacancies in chains of magnetic atoms on a superconductor. The starting point of the analysis is a finite perfect chain of magnetic atoms which contains a single vacancy or a dilute concentration of vacancy sites. This problem has, especially in the topologically nontrivial phase, close analogy to the problem of dilute magnetic impurities in a $s$-wave superconductor. In analogy to a subgap Shiba state bound to a single magnetic impurity, a single vacancy state gives rise to a localized low-energy state below the topological gap edge. This can be interpreted as two hybridized Majorana states at the weak link formed by the vacancy.  Also, the subgap spectrum of magnetic impurity systems and topological superconductors with vacancies is similar. In fact, the subgap DOS in the presence of single and dilute concentration of vacancies depicted in Figs.~\ref{ferrofour} (a), (c) and (d) would be qualitatively difficult to distinguish from the spectrum of an $s$-wave superconductor with weak concentration of magnetic impurities. However, at high vacancy concentrations there are qualitative differences between the two models.  In strongly disordered topological chains, lumped vacancy configurations give rise to domain walls of near zero energy Majorana end states. Thus, a topological chain at strong disorder will universally display a Griffiths effect manifesting as a singular density of states at the Fermi energy \cite{montrunich:2001:1,brouwer:2011:1,brouwer:2011:2}.  In our work we focused on finite chains with weak disorder, since we believe these to be closer to experimental interest.       

Experimentally the vacancy states could serve as a tool to verify the existence of the topological phase. Vacancy sites could be created by STM techniques that are used to probe the chains. Whether this is feasible or not depends on the chemistry of the adatoms and the surface properties of the employed materials. In principle, a single atom resolved manipulation is possible under suitable circumstances. Our results indicate that in the Shiba limit the single-vacancy subgap states are only present in the topologically nontrivial phase at physically relevant parameters. Creating a vacancy and identifying the resulting subgap state could serve as a smoking gun for the bulk topology. Furthermore, creating two vacancy states at different distances apart would reveal the hybrization of the vacancy states. In the dense chain the situation is not as clear cut as some trivial regions also support low-energy vacancy bound states.

The marked difference between the short-range model of densely-packed moments and the long-range Shiba models is that the latter exhibit fragile regions in the phase space where the single-vacancy bound state energy is tuned very close to the gap center. In these part of the topological phase diagram the gapped state is washed away by a low vacancy concentration. This effect is not correlated with the gap size and results in nucleation of gapless phase in the middle of the topological phase, splitting it to disconnected regions. In all models the deterioration of the topological phase at low vacancy concentration is driven by proliferation of the vacancy band, starting from the single vacancy energy and spreading to both directions until the gap of the clean system has been filled by a significant density of states. 

Finally, we also studied local fluctuations of the Shiba coupling that cause random shifts in the Shiba bound state energies. We studied how this type of disorder affects the topological band formation in the chains of magnetic atoms. In contrast to the vacancy disorder, this type of disorder only results in a proliferation of a gapless states nucleating from the phase boundaries. The single-Majorana phases of the Shiba chains with are found to be robust against moderate disorder of this type. 

\acknowledgments
The authors acknowledge the Academy of Finland and the Aalto University Center for Quantum Engineering for support.

\appendix
\numberwithin{equation}{section}

\section{Properties of clean systems}\label{appclean}

In this appendix, we will briefly discuss the clean limit of the three systems studied in the main text. We will derive their spectra, as well as the topological phase diagram from the $\mathbb{Z}_2$ invariant. Theoretical formulation of the helical and ferromagnetic Shiba chains follows Refs.~\cite{weststrom:2015:1} and \cite{poyhonen:2016:1}.
	
We note that under certain conditions the studied models  belong to the BDI symmetry class, and hence support $\mathbb{Z}$-valued invariants. In this work we are primarily concerned with the single Majorana phases and thus consider $\mathbb{Z}_2$ invariants. However, in the main text we also discuss the implications of disorder to the existence of double Majorana phases of ferromagnetic Shiba chains.

\subsection{Dense ferromagnetic chain}
	Starting from Eq.~\eqref{eq:dense:H} and considering a clean system, we obtain a BdG Hamiltonian of the form
	\begin{equation}
		H_k = (2t\cos k - \mu) \tau_z + B\sigma_z + 2\alpha_R\sin k \sigma_y \tau_z + \Delta \tau_x.
	\end{equation}
	This is now easily diagonalizable, yielding four energy bands
	\begin{widetext}
	\begin{equation}\label{cleanspectrum}
		E_k^2 = (2t\cos k - \mu)^2 + B^2 + 4\alpha_R^2\sin^2k + \Delta^2 \pm 2\sqrt{(2t\cos k - \mu)^2(B^2 +  4\alpha_R^2\sin^2k) + B^2\Delta^2}. 
	\end{equation}
	\end{widetext}
	From the gap closing conditions at $k = 0, \pi$, we can then derive analytical expressions for the boundaries of trivial and non-trivial $\mathbb{Z}_2$ phases: 
	\begin{equation}
		B^2 = (2t \pm \mu)^2 + \Delta^2.
	\end{equation}
The phase boundaries in Fig.~\ref{ferrotwo} are obtained from this result. 	

\subsection{Ferromagnetic Shiba chain}

	For a Shiba chain with magnetic impurities placed at positions $\vec r_i$ and Rashba SOC, the BdG Hamiltonian reads
	\begin{equation}\label{app:eq:shiba}
	\begin{split}
		H = \left(\frac{\vec p^2}{2m}-\mu + \alpha_R(p_y \sigma_x - p_x \sigma_y)\right)\tau_z + \Delta \tau_x \\
		 - J\sum_{i} (\vec S_i \cdot \boldsymbol{\sigma})\delta(\vec r - \vec r_i),
	\end{split}
	\end{equation}
	where $\mu,\ \alpha_R$, and $\Delta$ are defined as previously, $m$ is the mass of the electron, $J$ is the coupling strength between the magnetic impurities and the electrons, and finally $\vec S_i$ is the magnetic moment at $\vec r_i$. We have also defined $\boldsymbol{\sigma} \equiv (\sigma_x, \sigma_y, \sigma_z)$.
	
	Assuming that the impurities form a 1D chain with a lattice constant $a$, it is possible to reduce the subgap spectral problem to the form
	\begin{equation}\label{app:eq:rasbhaJE}
		\Psi(x_i) = \sum_j \alpha J_E(x_{ij})\Psi(x_j),
	\end{equation}
	where $x_{ij} = x_i-x_j=(i-j)a$ and the coupling $J_E$, which is essentially given by the Green's function of the bulk and the magnetic texture \cite{brydon:2015:1}. Furthermore, we have also introduced the dimensionless Shiba coupling $\alpha = \pi JS\nu_0$, where $\nu_0$ is the density of states at the Fermi energy. 

	For the ferromagnetic case, we have that $\vec S_i = S \hat{\vec z}$, where $\hat{\vec z}$ is the unit vector in the $z$ direction. Following the steps in Ref.~\cite{poyhonen:2016:1}, we can reformulate the subgap spectrum of $N$ magnetic sites into a NLEVP of the form $\tilde{G}^{-1}(E) \Psi(E) = 0$, where
	\begin{equation}\label{app:eq:rashbaNLEVP}
		\tilde{G}^{-1}(E) =\begin{pmatrix}
A\lambda^2-\tfrac{\lambda}{\alpha} & B\lambda & C\lambda^2 & -\lambda D\\
-B\lambda & -A + \tfrac{\lambda}{\alpha} & -\lambda D & C\\
-C\lambda^2 & -\lambda D & A\lambda^2 + \tfrac{\lambda}{\alpha} & -B\lambda\\
-\lambda D & -C & B \lambda & - A -\tfrac{\lambda}{\alpha}
\end{pmatrix},\\
	\end{equation}
 $\lambda = (\Delta + E)/\sqrt{\Delta^2 - E^2}$, and $A,\ B,\ C$, and $D$ are $N \times N$ matrices
	\begin{equation}
		\begin{split}
A_{ij} &= -\frac{1}{2m}(I_3^-(x_{ij}) + I_3^+(x_{ij})) + \delta_{ij}\\
B_{ij} &= - \frac{i}{2m}(I_2^-(x_{ij}) - I_2^+(x_{ij}))\\
C_{ij} &= -\frac{i}{2m}(I_4^-(x_{ij}) - I_4^+(x_{ij}))\\
D_{ij} &= - \frac{1}{2m}(I_1^-(x_{ij}) + I_1^+(x_{ij}))
\end{split}
	\end{equation}	
	 defined using
	 \begin{widetext}
	\begin{equation}
	\begin{split}
		I^\pm_1(x) &= N_\pm\IM\left[J_0((k_{F,\pm} + i\xi_E^{-1})|x|))+ i H_0((k_{F,\pm} + i\xi_E^{-1})|x|))\right]\\
		I^\pm_2(x) &= -iN_\pm\sgn(x)\RE\left[iJ_1((k_{F,\pm} + i\xi_E^{-1})|x|))+ H_{-1}((k_{F,\pm} + i\xi_E^{-1})|x|))\right]\\
		I^\pm_3(x) &= -N_\pm\RE\left[J_0((k_{F,\pm} + i\xi_E^{-1})|x|))+ i H_0((k_{F,\pm} + i\xi_E^{-1})|x|))\right]\\
		I^\pm_4(x) &= -i N_\pm\sgn(x)\IM\left[iJ_1((k_{F,\pm} + i\xi_E^{-1})|x|))+ H_{-1}((k_{F,\pm} + i\xi_E^{-1})|x|))\right],
	\end{split}
	\end{equation}
	\end{widetext}
	where $J_\nu$ and $H_\nu$ are the Bessel function of the first kind  and Struve function, respectively. We have also introduced the shorthands $N_\pm = 1 \mp \zeta/\sqrt{1+\zeta^2}$, $k_{F,\pm} = k_F(\sqrt{1+\zeta^2} \mp \zeta)$, and $\xi_E = v_F/\sqrt{\Delta^2 - E^2} = \xi_0/\sqrt{1 - E^2/\Delta^2}$. In these expressions, we have used the dimensionless Rashba coupling $\zeta = m\alpha_R/k_F$, along with the Fermi wavenumber $k_F$ and velocity $v_F$. The quantity $\xi_0  = v_F/\Delta$ is the superconducting coherence length. In the limit of infinite coherence length, all energy dependence except for the one in $\lambda$ vanishes, giving us a polynomial eigenvalue problem in $\lambda$. Polynomial NLEVPs of order $n$ and dimension $N$ can be written as generalized linear eigenvalue problems of size $nN\times nN$. As discussed in Ref.~\cite{poyhonen:2016:1}, it turns out that setting $\xi_E \to \xi_0$ in the block matrices is an excellent approximation which enables us to treat the NLEVP \eqref{app:eq:rashbaNLEVP} as a second order polynomial eigenvalue problem in $\lambda$. 
	
	Since the system is once again translation invariant, we can block diagonalize the matrix in momentum space and obtain an analytical expression for the spectrum. In terms of the Fourier transforms of the individual $N\times N$ matrices, we get
	\begin{widetext}
	\begin{equation}
		E_k^2= \Delta^2\frac{(A_k^2 + B_k^2+C_k^2+D_k^2-1/\alpha)^2-4(A_kB_k+C_kD_k)^2}{(A_k^2 + B_k^2+C_k^2+D_k^2-1/\alpha)^2-4(A_kB_k+C_kD_k)^2+4(A_k^2+C_k^2)}.
	\end{equation}
	\end{widetext}
	The topological $\mathbb{Z}_2$ phase boundaries can be extract from Eq.~\eqref{app:eq:rashbaNLEVP} by setting $E = 0$ together with $ka = 0, \pi$ and requiring that $\mathrm{det}\, \tilde{G}^{-1}=0$. This yields an equation of the form
	\begin{equation}
		\alpha_{0,\pi} = \frac{1}{\sqrt{A_k^2+D_k^2}}\bigg|_{ka=0,\pi}.
	\end{equation}
	Above procedure results in simple exact expressions for the $\mathbb{Z}_2$  phase boundaries employed in the main text.

\subsection{Helical Shiba chain}

	In the helical case, we start from \eqref{app:eq:shiba} by setting $\alpha_R = 0$, and $\vec S_j = S(\cos\phi_j\sin\theta, \sin\phi_j\sin\theta, \cos\theta)$, where $\phi_i$ is directly proportional to $x_j$. In other words, $\phi_j = 2k_H x_j=2jk_Ha$, where the helical wavenumber $k_H$ determines the pitch and $a$ is the lattice constant. We now also assume that the superconducting bulk is three dimensional. 
	
	Following the steps in Ref.~\cite{weststrom:2015:1}, we can -- similarly to the ferromagnetic case -- transform this problem into a NLEVP $\tilde{G}^{-1}(E) \Psi(E) = 0$, where
	\begin{equation}\label{app:eq:helixNLEVP}
		\begin{split}		
		\tilde{G}^{-1}=\begin{pmatrix}
			\lambda^2h^{\uparrow\uparrow}-\frac{\lambda}{\alpha} & -\lambda d^{\uparrow\downarrow} & -\lambda^2h^{\uparrow\downarrow} & \lambda d^{\uparrow\uparrow}\\
			-\lambda d^{\downarrow\uparrow} & -h^{\downarrow\downarrow}+ \frac{\lambda}{\alpha} & \lambda d^{\downarrow\downarrow} & h^{\downarrow\uparrow}\\
			-\lambda^2h^{\downarrow\uparrow} & \lambda d^{\downarrow\downarrow} & \lambda^2h^{\downarrow\downarrow}+ \frac{\lambda}{\alpha} & -\lambda d^{\downarrow\uparrow}\\
			\lambda d^{\uparrow\uparrow} & h^{\uparrow\downarrow} & -\lambda d^{\uparrow\downarrow} & -h^{\uparrow\uparrow} - \frac{\lambda}{\alpha}\\
		\end{pmatrix}
		.
	\end{split}
	\end{equation}
	The subblock matrices are defined according to
	\begin{equation}
		\begin{split}
			h^{\sigma\sigma^\prime}_{ij} &\equiv \delta_{ij}\delta_{\sigma\sigma'} +  \Gamma_{ij}\sin (k_F|x_{ij}|)\bra{\sigma}\sigma^\prime\rangle_{ij}\\ d^{\sigma\sigma^\prime}_{ij} &\equiv \Gamma_{ij}\cos (k_F|x_{ij}|)\bra{\sigma}\sigma^\prime\rangle_{ij},
		\end{split}
	\end{equation}
	where
	\begin{equation}
	\begin{split}
		&\Gamma_{i\neq j} \equiv \frac{1}{k_F|x_{ij}|}e^{-\frac{|x_{ij}|}{\xi_E}},\quad \Gamma_{ii} \equiv 0\\
		&\brakets{\uparrow}{\uparrow}_{ij} = \brakets{\downarrow}{\downarrow}_{ij}^* = \cos^2\frac{\theta}{2}e^{ik_Hx_{ij}} + \sin^2\frac{\theta}{2}e^{-ik_Hx_{ij}}\\
		&\brakets{\uparrow}{\downarrow}_{ij} = \brakets{\downarrow}{\uparrow}_{ij} = i\sin\theta\sin k_Hx_{ij}.
	\end{split}
	\end{equation}
	After performing a Fourier transformation of the block matrices, the spectrum can be found by setting the determinant of the matrix in Eq.~\eqref{app:eq:helixNLEVP} to zero. This procedure yields the energy bands
	\begin{equation}
		E_{\beta\gamma}(k) = \Delta\frac{\lambda_{\beta\gamma}(k)^2-1}{\lambda_{\beta\gamma}(k)^2+1},
	\end{equation}
	where $\beta$,$\gamma = \pm 1$ are independent signs, and
	\begin{equation}
	\begin{split}
		&\lambda_{\beta\gamma}(k) = \beta \frac{\sqrt{B^2-4AC-8A^2}}{4A}  - \frac{B}{4A} \\
		&+\frac{\gamma}{2}\sqrt{\frac{B^2}{2A^2} + \beta\frac{8B+4\frac{BC}{A} - \frac{B^3}{A^2}}{2\sqrt{B^2-4AC-8A^2}} - \frac{C}{A} + 2},
	\end{split}
	\end{equation}
	where we have defined the functions
	\begin{equation}
		\begin{split}
			A = &\alpha^2\left[(h_{k}^{\uparrow\downarrow})^2-h_k^{\uparrow\uparrow}h_{-k}^{\uparrow\uparrow}\right]\\
			B = &\alpha^3\left[h_k^{\uparrow\uparrow}(d_{-k}^{\uparrow\uparrow})^2{-}h_{-k}^{\uparrow\uparrow}(d_k^{\uparrow\uparrow})^2+2d_{k}^{\uparrow\downarrow}h_{k}^{\uparrow\downarrow}(d_k^{\uparrow\uparrow}{-}d_{-k}^{\uparrow\uparrow})\right]\\
			& + \alpha^3(h_{-k}^{\uparrow\uparrow}-h_k^{\uparrow\uparrow})\left[h_k^{\uparrow\uparrow}h_{-k}^{\uparrow\uparrow}+(d_{k}^{\uparrow\downarrow})^2 -(h_{k}^{\uparrow\downarrow})^2+\alpha^{-2}\right]\\ 
			C = &\alpha^4 \left[(d_{k}^{\uparrow\downarrow})^2-(h_{k}^{\uparrow\downarrow})^2+h_k^{\uparrow\uparrow}h_{-k}^{\uparrow\uparrow}-d_k^{\uparrow\uparrow}d_{-k}^{\uparrow\uparrow}\right]^2\\
			&+\alpha^4\left[2d_{k}^{\uparrow\downarrow}h_{k}^{\uparrow\downarrow}-h_k^{\uparrow\uparrow}d_{-k}^{\uparrow\uparrow}-d_k^{\uparrow\uparrow}h_{-k}^{\uparrow\uparrow}\right]^2 + 1\\
			&+\alpha^2\left[2(d_{k}^{\uparrow\downarrow})^2-(d_k^{\uparrow\uparrow})^2-(d_{-k}^{\uparrow\uparrow})^2 - (h_k^{\uparrow\uparrow}-h_{-k}^{\uparrow\uparrow})^2\right]
		\end{split}
	\end{equation}
	
	The topological phase boundaries are derived using the same approach as before: set $E=0$ everywhere in the Fourier transformed \eqref{app:eq:helixNLEVP}, set $ka = 0, \pi$, and require $\mathrm{det}\, \tilde{G}^{-1}=0$. This gives us the phase boundaries in a compact form
	\begin{equation}
		\alpha_{0,\pi} = \frac{1}{\sqrt{(h_k^{\uparrow\uparrow})^2+(d_k^{\uparrow\uparrow})^2}}\Bigg\vert_{ka=0,\pi}.
	\end{equation}
	For a planar helix, this gives the complete phase diagram for the $\mathbb{Z}_2$ invariant, but as soon as we move away from $\theta = \pi/2$, gapless regions emerge. These cannot be accounted for using this procedure since the gapless regions have gap closings away from $ka = 0,\pi$. However, the above relation provides a complete $\mathbb Z_2$ description of a planar helix.

\section{Topological Hamiltonians for Shiba chains}\label{topoH}

\subsection{Interpretation of $\tilde{G}^{-1}(E)$}

We begin by elucidating the interpretation of the matrix $\tilde{G}^{-1}$ which appears as a fundamental object in the low energy theory of Shiba chains. 

\begin{equation}
(E-H_0)\Psi \equiv G_0^{-1}\Psi = \sum_jV_j\delta(\vec r - \vec r_j)\Psi,
\end{equation}
where $G_0$ is the unperturbed Green's function of the underlying superconductor and $V_j=J(\vec S_j \cdot \boldsymbol{\sigma})$. This can be further written as 
\begin{equation}
\Psi(\vec r) = \sum_j G_0(\vec r - \vec r_j)V_j \Psi(\vec r_j).
\end{equation}
By restricting the position vector to the positions of magnetic atoms and introducing notation $\Psi(\vec r_i)=\Psi_i$ we obtain a closed eigenvalue problem $\Psi_i = \sum_j G_0(\vec r_{ij})V_j \Psi_j$. Regarding $G_0(\vec r_{ij})$ and $V_j$ as matrices in the site indices, we can write the eigenvalue problem in more abstract matrix notation as
\begin{equation}
\left(\mathbb{I} -  G_0V\right)\Psi = G_0\left(G_0^{-1} -  V \right)\Psi = 0
\end{equation}
or equivalently as
\begin{equation}
G_0G^{-1}\Psi \equiv \tilde{G}^{-1}\Psi = 0.
\end{equation}
Hence we see that the matrix $\tilde{G}^{-1}$ can be viewed as a product of the Green's function of the unperturbed superconductor and the inverse Green's function of the full system restricted to the magnetic sites.  The spectrum of the chain can be found as poles for the Green's function $\mathrm{det}\, G^{-1}(E)=0$. Since $G_0$ does not have subgap poles or zeros, the subgap spectrum can equivalently be obtained from $\mathrm{det}\, \tilde{G}^{-1}(E)=0$. Since also the subgap kernel of $\tilde{G}^{-1}(E)$ and $G^{-1}(E)$ coincide, the difference between them is largely immaterial.

\subsection{Relation between $\tilde H$ and $\tilde{G}^{-1}(E)$}

While the NLEVP matrices in Eqs.~\eqref{app:eq:rashbaNLEVP},\eqref{app:eq:helixNLEVP} contain all the information about the subgap energy bands and their topology, the straightforward solution of the spectral problem is rather resource intensive and a more expedient method is hence desirable. Also, typically the topological properties are extracted from a 1D Hamiltonian of the system which is not the fundamental object in the low-energy effective theory. In clean systems, it is possible to obtain the winding number from the NLEVP matrix in reciprocal space as outlined in Ref.~\cite{poyhonen:2016:1}. However, in disordered systems momentum is not a good quantum number and alternative methods are required. In cases where only the parity of the topology is relevant, the $\mathbb{Z}_2$ invariant can be obtained from the Pfaffian of the Hamiltonian in a suitable basis, avoiding the need for an explicit solution of the eigenvalues and vectors. Therefore it is desirable to find a similar approach. 

For a given NLEVP
 \begin{equation}\label{Gspec}
\tilde{G}^{-1}(E) \Psi(E) = 0,
\end{equation}
we can define a topological Hamiltonian $\tilde H = \tilde{G}^{-1}(0)$ with an auxiliary spectral problem as
\begin{equation}\label{Hspec}
\tilde H \Psi_{\tilde H}(E) = E\Psi_{\tilde H}(E).
\end{equation}
We stress that $\tilde H$ should \textit{not} be employed in finding the spectrum of the system. However, a closer look reveals that the topological properties of models \eqref{Gspec} and \eqref{Hspec} can be obtained from one another: in the limit $E\to 0$, we have $\Psi(E) \to \Psi_{\tilde H}(E)$ and $\tilde{G}^{-1}(E) \to \tilde H$. In particular, this means that the gap closings of the two systems \emph{coincide}, and that their zero energy solutions are \emph{exactly} identical. Further, the Hamiltonian $\tilde H$ inherits the symmetries of the original system. Consequently, we can obtain the $\mathbb{Z}_2$ topological phase diagram of the original system by considering that of the effective system introduced here; as the Pfaffian only changes sign at gap closings, and a negative sign corresponds directly to zero energy modes \cite{kitaev:2001:1}, the phase diagrams will also coincide.

To regard  $\tilde H$ as a Hamiltonian, it must be Hermitian. This is true in the clean limit and for vacancy disorder, where the NLEVP matrices in Eqs.~\eqref{app:eq:helixNLEVP} and \eqref{app:eq:rashbaNLEVP} are Hermitian. However, the case of local $\alpha$ disorder requires a slight reformulation of the problem. Going back to Eq.~\eqref{app:eq:rasbhaJE} but allowing for $\alpha$ to vary locally, we have
\begin{equation}
\Psi(x_i) = \sum_j \alpha_j J_E(x_{ij}) \Psi(x_j).
\end{equation}
Multiplying both sides by $\alpha_i \neq 0$ and then going through the steps to obtain the NLEVP, we find in the ferromagnetic case
\begin{equation}\label{eq:nlevprime}
\begin{pmatrix}
a^\prime\lambda^2-\beta\lambda & b^\prime\lambda & c^\prime\lambda^2 & -\lambda d^\prime\\
-b^\prime\lambda & \beta\lambda - a^\prime & -\lambda d^\prime & c^\prime\\
-c^\prime\lambda^2 & -\lambda d^\prime & a^\prime\lambda^2 + \beta\lambda & -b\lambda\\
-\lambda d^\prime & -c^\prime & b^\prime \lambda & -\beta\lambda - a^\prime
\end{pmatrix}
\Psi = 0,
\end{equation}
where we have introduced the matrix $\beta_{ij} = \alpha_i\delta_{ij}$, and the prime on the other submatrices indicate $a^\prime_{ij} = \alpha_i\alpha_j a_{ij}$. In the limit $E\to 0$ this lets us define a Hamiltonian $\tilde H$ as outlined previously in this chapter; as the  resultant Hamiltonian is a real matrix in the BDI symmetry class, it can conveniently be antisymmetrized through multiplication with the particle-hole symmetry operator, as was done in the dense chain. This allows a direct calculation of the Pfaffian, yielding the phase diagram of the system. In the helical model the process is somewhat more involved: while a Hermitian $\tilde H$ can be similarly obtained, it is a complex Hamiltonian in the symmetry class D, preventing simple antisymmetrization by use of the PHS operator. However, in the planar limit, $\theta = \pi/2$, it also in symmetry class BDI. Consequently the effective Hamiltonian in that model can be made real by a unitary transformation $\tilde H \to U^\dagger\tilde H U$ with $U = \exp(i\frac{\pi}{4}\tau_z\sigma_z)$, and subsequently antisymmetrized to allow calculation of the Pfaffian. As seen in Ref.~\cite{weststrom:2015:1}, the $\mathbb Z_2$ phase boundaries of the helical model are independent of the angle $\theta$, and hence the phase diagram thus obtained is correct for general parameters with the caveat that the boundaries of gapless phase is not captured by the approach.

\bibliographystyle{h-physrev}

\end{document}